# The Complexity of Quantified Constraint Satisfaction: Collapsibility, Sink Algebras, and the Three-Element Case

Hubie Chen
Departament de Tecnologia
Universitat Pompeu Fabra
Barcelona, Spain
*hubie.chen@upf.edu*

**Abstract**

The constraint satisfaction probem (CSP) is a well-acknowledged framework in which many combinatorial search problems can be naturally formulated. The CSP may be viewed as the problem of deciding the truth of a logical sentence consisting of a conjunction of constraints, in front of which all variables are existentially quantified. The quantified constraint satisfaction problem (QCSP) is the generalization of the CSP where universal quantification is permitted in addition to existential quantification. The general intractability of these problems has motivated research studying the complexity of these problems under a restricted *constraint language*, which is a set of relations that can be used to express constraints.

This paper introduces collapsibility, a technique for deriving positive complexity results on the QCSP. In particular, this technique allows one to show that, for a particular constraint language, the QCSP reduces to the CSP. We show that collapsibility applies to three known tractable cases of the QCSP that were originally studied using disparate proof techniques in different decades: QUANTIFIED 2-SAT (Aspvall, Plass, and Tarjan 1979), QUANTIFIED HORN-SAT (Karpinski, Kleine Büning, and Schmitt 1987), and QUANTIFIED AFFINE-SAT (Creignou, Khanna, and Sudan 2001). This reconciles and reveals common structure among these cases, which are describable by constraint languages over a two-element domain. In addition to unifying these known tractable cases, we study constraint languages over domains of larger size.

## 1 Introduction

**Background.** The *constraint satisfaction problem (CSP)* is a general framework in which many combinatorial search problems can be naturally formulated. An instance of the CSP consists of a set of variables, a domain, and a set of *constraints* on the variables, where a constraint is a pair consisting of a tuple $(v_1, \ldots, v_k)$ of variables and a relation $R \subseteq D^k$ over the domain $D$ that specifies allowed values for the variable tuple. The question is to decide whether or not there exists an assignment to the variables satisfying all of the constraints. Examples of problems that fall into this framework include boolean satisfiability problems [30, 12], graph homomorphism problems [19, 20], and the problem of solving a system of equations over an algebraic structure [27]. The CSP can be equivalently formulated as the problem of deciding if there is a homomorphism between two relational structures [18], as well as the database problems of conjunctive-query containment and evaluation [26].

In its general formulation, the CSP is NP-complete; this intractability, coupled with the ubiquity of the CSP, has given rise to a far-reaching research program that aims to identify restricted cases of the CSP that are polynomial-time tractable. One of the principal directions within this program is the study of the CSP where the set of relations permitted in constraints, called the *constraint language*, is restricted. This direction has its origins in a 1978 paper of Schaefer [30], who gave a classification theorem showing that all constraint languages over a two-element domain give rise to a case of the CSP that is either polynomial-time tractable or NP-complete. The non-trivial tractable cases given by this result all readily reduce to one of the following three problems: 2-SAT, where each constraint is equivalent to a clause of length 2; HORN-SAT, where each constraint is equivalent to a propositional Horn clause; or AFFINE-SAT, where each constraint is an equation over the two-element field. In the nineties, an approach to studying the complexity of constraint languages based on concepts and tools from universal algebra was introduced [23]. A key idea underlying this approach is to associate, to each constraint language, an algebra whose operations are the *polymorphisms* of the constraint language; roughly speaking, an operation is a polymorphism of a constraint language $\Gamma$ if each relation of $\Gamma$ is closed under the operation. This algebra is then used to derive information about the constraint language. One celebrated achievement of this algebraic viewpoint is the CSP complexity classification of constraint languages over a three-element domain, due to Bulatov [9]. See [14, 8, 5, 7, 16, 6, 25] and the references therein for more examples of work along these lines.

The present work is concerned with the *quantified constraint satisfaction problem (QCSP)*. Viewing CSP as the problem of deciding a logical sentence consisting of a conjunction of constraints and a quantifier prefix in which all variables are existentially quantified, the QCSP can be defined as the generalization of the CSP where universal quantification is permitted in addition to existential quantification. As is well-known, the extra expressiveness of the QCSP comes at the cost of complexity: the QCSP is in general PSPACE-complete. Indeed, the QUANTIFIED BOOLEAN FORMULA (QBF) problem, of which the QCSP is a generalization, was historically one of the first problems recognized to be PSPACE-complete [31], and is now a prototypical example of a PSPACE-complete problem.

A classification theorem describing the QCSP complexity of constraint languages over a two-element domain has been established [30, 13, 12]. The polynomial-time tractable cases of the QCSP given by this classification correspond to the problems QUANTIFIED 2-SAT, QUANTIFIED HORN-SAT, and QUANTIFIED AFFINE-SAT. These are precisely the quantified generalizations of the discussed tractable cases of Schaefer's CSP theorem. Although the tractability of all three of these problems was claimed without proof by Schaefer [30], the first published proofs we are aware of for these problems were given in different decades, and proved using disparate proof techniques: Aspvall, Plass, and Tarjan [1] proved the tractability of QUANTIFIED 2-SAT in 1979 by giving an algorithm which analyses a directed graph, called the *implication graph*, induced by a problem instance; Karpinski, Kleine Büning, and Schmitt [24] proved the tractability of QUANTIFIED HORN-SAT in 1987 by studying a generalized form of unit resolution (see [10] for subsequent work on this problem), and Creignou, Khanna, and Sudan [12] proved the tractability of QUANTIFIED AFFINE-SAT in 2001 by giving an algorithm that iteratively eliminates the innermost quantified variable.

**Contributions.** In this article, we introduce *collapsibility*, a technique for deriving positive complexity results on the QCSP, which is based on the algebraic approach to studying constraint languages. This technique allows one to show, for certain constraint languages, that the QCSP over the constraint language polynomial-time reduces to the CSP over the constraint language, via a simple reduction. Thus, this technique, when applied to a constraint language under which the CSP is known to be tractable, implies the tractability of the QCSP under the constraint language. *We show that this technique applies to all three of the aforementioned tractable cases of the QCSP over a two-element domain.* This reconciles and reveals common structure among these three cases, which, as discussed, were originally studied using strikingly disparate techniques.

In addition to unifying known QCSP tractability results, the technique of collapsibility allows us to make significant progress towards understanding the complexity of the QCSP in domains of arbitrary size. For example, we are able to demonstrate the QCSP tractability of the large classes of constraint languages studied, in the CSP setting, by Jeavons, Cohen, and Cooper [22] and Bulatov and Dalmau [6]. We are also able to classify the *conservative constraint languages* giving rise to a tractable QCSP; a conservative constraint language is a constraint language containing all unary relations, and such constraint languages were studied by Bulatov [5] in the CSP.

Based on the notion of collapsibility, we identify a class of algebras that we call *sink algebras*, and show that any constraint language whose algebra "excludes" sink algebras (in a manner made precise) and does not obey a known sufficient condition for CSP intractability, is amenable to our collapsibility technique. We analyze three-element sink algebras and show that they all contain a particular semilattice operation. This in turn allows us to give a classification for the three-element case up to a "forbidden polymorphism": for the constraint languages over a three-element domain not having the semilattice operation as polymorphism, we provide a description of exactly which give rise to a tractable QCSP.

Overall, the techniques of this article involve an interplay among the areas of complexity theory, algebra, and logic.

**Other related work.** The other papers on the QCSP and constraint languages on domains of arbitrary size that we are aware of are Börner et al. [3] and Martin and Madelaine [28]. The results of Börner et al. [3] include the identification of a Galois connection relevant to the QCSP and two tractable cases of the QCSP. Our collapsibility technique applies to the tractable cases they give, which are defined in terms of Mal'tsev polymorphisms and dual discriminator polymorphisms. Martin and Madelaine [28] study constraint languages consisting of a single binary relation that is symmetric and reflexive, that is, an undirected graph; they obtain both tractability and intractability results for such constraint languages.

**Organization.** This article is organized as follows. In Section 2, we present the terminology, notation, and background concepts to be used throughout the paper. In Section 3, we give the basic definitions and results that underlie our collapsibility technique. In particular, we define what it means for a constraint language to be *collapsible*, and show that the QCSP over a collapsible constraint language reduces to the CSP over the same constraint language. Section 4 presents a theorem that can be used to prove the collapsibility of constraint languages, and gives example applications of the theorem. In Section 5, we develop algebraic machinery for demonstrating collapsibility results, and illustrate our ideas using examples. Section 6 defines and studies *sink algebras*. In Section 7, we analyze three-element sink algebras, showing that any such algebra must have a particular structure; this permits us to give our classification of constraint languages over a three-element domain.

## 2 Preliminaries

We use $[n]$ to denote the set containing the first $n$ positive integers, $\{1, \ldots, n\}$. When $f : A^k \to A$ is an operation on a set $A$ and $B_1, \ldots, B_k \subseteq A$ are subsets of $A$, we use $f(B_1, \ldots, B_k)$ to denote the set $\{f(b_1, \ldots, b_k) : b_1 \in B_1, \ldots, b_k \in B_k\}$. When $f : A^k \to A$ is an operation on a set $A$ and $B \subseteq A$, we use $f|_B$ to denote the restriction of $f$ to $B^k$, and when $F$ is a set of functions $f : A^k \to A$, we use $F|_B$ to denote $\{f|_B : f \in F\}$.

### 2.1 Quantified Constraint Satisfaction

We now describe the basic terminology of quantified constraint satisfaction to be used throughout the paper.

**Definition 2.1** *Let $A$ be a set. A* relation *over the set $A$ is a subset of $A^k$ for some integer $k \geq 1$, called the arity of the relation. A* constraint *over the set $A$ is a formula of the form $R(w_1, \ldots, w_k)$ where $R$ is a* *relation over A of arity k (viewed as a predicate), and each $w_i$ is either a variable or a* constant *(an element of A). A* constraint language *is a set of relations, all of which are all over the same domain.*

Note that in this paper, we always permit constants to appear in constraints. Clearly, our positive complexity results will apply to constraints not containing any constants, since such constraints are a subclass of the constraints we allow here.

**Definition 2.2** *A* quantified constraint formula *is a formula of the form*

$$Q_1 v_1 \ldots Q_m v_m \mathcal{C}$$

*having an associated set A called the domain, where:*

- *for all $i \in [m]$, $Q_i$ is a quantifier from the set $\{\forall, \exists\}$ and $v_i$ is a variable;*
- *the variables $\{v_1, \ldots, v_m\}$ are assumed to be pairwise distinct, and,*
- *$\mathcal{C}$ is a finite conjunction of constraints over A having variables from $\{v_1, \ldots, v_m\}$.*

*A quantified constraint formula is said to be over a constraint language $\Gamma$ if each of its constraints has relation from $\Gamma$.*

Truth of a quantified constraint formula is defined as in first-order logic. Note that the quantification of the variables is understood to be over the domain $A$ of the formula. We will generally use $A$ to denote the domain of a quantified constraint formula. We assume that all domains of quantified constraint formulas are finite.

The QCSP can now be defined as the problem of deciding, given a quantified constraint formula, whether or not it is true. We are interested in the following parameterized version of the QCSP.

**Definition 2.3** *Let $\Gamma$ be a constraint language. The decision problem $\mathsf{QCSP_c}(\Gamma)$ is to decide, given as input a quantified constraint formula over $\Gamma$, whether or not it is true.*

We will also discuss and use the following parameterized version of the CSP.

**Definition 2.4** *The decision problem $\mathsf{CSP_c}(\Gamma)$ is the restriction of $\mathsf{QCSP_c}(\Gamma)$ to quantified constraint formulas having only existential quantifiers.*

In the previous two definitions, we use the subscript c to emphasize that constants are permitted in constraints.

We now review a characterization of truth for quantified constraint formulas, which will be used throughout this paper. This characterization comes from the concept of Skolemization [17], and conceives of a quantified constraint formula as a game between two players: a *universal player* that sets the universally quantified variables, and an *existential player* that sets the existentially quantified variables. Variables are set in the order dictated by the quantifier prefix, and the existential player is said to win if, after the variables have been set, the conjunction of constraints is true. The formula is true if and only if the existential player can always win, no matter how the universal player sets the universally quantified variables.

We now formalize this viewpoint. When $\Phi$ is a quantified constraint formula, let $V^\Phi$ denote the variables of $\Phi$, let $E^\Phi$ denote the existentially quantified variables of $\Phi$, let $U^\Phi$ denote the universally quantified variables of $\Phi$, and for each $x \in E^\Phi$, let $U_x^\Phi$ denote the variables in $U^\Phi$ that come before $x$ in the quantifier prefix of $\Phi$. Let $[B \to A]$ denote the set of functions mapping from $B$ to $A$.

**Definition 2.5** *A* strategy *for a quantified constraint formula $\Phi$ is a sequence of partial functions*

$$\sigma = \{\sigma_x : [U_x^\Phi \to A] \to A\}_{x \in E^\Phi}.$$

That is, a strategy has a mapping $\sigma_x$ for each existentially quantified variable $x \in E^\Phi$, which tells how to set the variable $x$ in response to an assignment to the universal variables coming before $x$. Let $\tau : U^\Phi \to A$ be an assignment to the universal variables. We define $\langle \sigma, \tau \rangle$ to be the mapping from $V^\Phi$ to $A$ such that $\langle \sigma, \tau \rangle(v) = \tau(v)$ for all $v \in U^\Phi$, and $\langle \sigma, \tau \rangle(x) = \sigma_x(\tau|_{U_x^\Phi})$ for all $x \in E^\Phi$. The mapping $\langle \sigma, \tau \rangle$ is undefined if $\sigma_x(\tau|_{U_x^\Phi})$ is not defined for all $x \in E^\Phi$. The intuitive point here is that a strategy $\sigma$ along with an assignment $\tau$ to the universally quantified variables naturally yields an assignment $\langle \sigma, \tau \rangle$ to all of the variables, so long as the mappings $\sigma_x$ are defined at the relevant points.

We have the following characterization of truth for quantified constraint formulas.

**Fact 2.6** *A quantified constraint formula $\Phi$ is true if and only if there exists a strategy $\sigma$ for $\Phi$ such that for all mappings $\tau : U^\Phi \to A$, the assignment $\langle \sigma, \tau \rangle$ is defined and satisfies the constraints of $\Phi$.*

Note that a strategy satisfying the condition of Fact 2.6 must consist only of total functions. We have defined a strategy to be a sequence of *partial* functions as we will be interested in strategies $\sigma$ that need not yield an assignment $\langle \sigma, \tau \rangle$ for all $\tau$.

## 2.2 Algebra

This subsection presents the algebraic background used in this paper; for more information, we refer the reader to the books [29, 32]. We begin by defining the notion of polymorphism. This notion will be used in this subsection to define further algebraic notions, and, as explained in the next subsection, will also be used throughout the paper to study the complexity of constraint languages.

**Definition 2.7** *An operation $f : A^k \to A$ is a* polymorphism *of a relation $R \subseteq A^m$ if for any choice of $k$ tuples $(t_{11}, \ldots, t_{1m}), \ldots, (t_{k1}, \ldots, t_{km}) \in R$, the tuple $(f(t_{11}, \ldots, t_{k1}), \ldots, f(t_{1m}, \ldots, t_{km}))$ is in $R$. That is, applying $f$ coordinate-wise to any $k$ tuples in $R$ yields another tuple in $R$. An operation $f$ is a polymorphism of a constraint language $\Gamma$ if $f$ is a polymorphism of all relations $R \in \Gamma$. When $f$ is a polymorphism of a relation $R$ (respectively, a constraint language $\Gamma$), we also say that $R$ (respectively, $\Gamma$) is* invariant *under $f$.*

We now introduce the notion of a *clone*.

**Definition 2.8** *An operation $f : A^k \to A$ is a* projection *if there exists $i \in [k]$ such that $f(a_1, \ldots, a_k) = a_i$ for all $a_1, \ldots, a_k \in A$. When $f : A^n \to A$ is an arity $n$ operation and $f_1, \ldots, f_n : A^m \to A$ are arity $m$ operations, the composition of $f$ with $f_1, \ldots, f_n$ is defined to be the arity $m$ operation $g : A^m \to A$ such that $g(a_1, \ldots, a_m) = f(f_1(a_1, \ldots, a_m), \ldots, f_n(a_1, \ldots, a_m))$ for all $a_1, \ldots, a_m \in A$.*

**Definition 2.9** *A* clone *on a set $A$ is a set of finitary operations that contains all projections and is closed under composition.*

It is known that the set of polymorphisms of a constraint language is always a clone.
Next, we define the basic notion of an *algebra*.

**Definition 2.10** *An* algebra *is a pair $\mathbb{A} = (A, F)$ where $A$ is a non-empty set called the* universe, *and $F$ is a set of finitary operations on $A$.*

We say that an algebra is *non-trivial* if its universe does not have size one. An algebra is *finite* if its universe is of finite size.

**Definition 2.11** *An operation $f$ on the set $A$ is* idempotent *if $f(a,\ldots,a) = a$ for all $a \in A$. An algebra $(A, F)$ is* idempotent *if all operations in $F$ are idempotent.*

Note that, in this paper, we are concerned almost exclusively with finite idempotent algebras.

**Definition 2.12** *The* term operations *of an algebra $(A, F)$ are the operations in the clone generated by $F$.*

We now present two standard means of constructing new algebras from an existing algebra. The first is the construction of a *subalgebra* from an algebra.

**Definition 2.13** *Let $\mathbb{A} = (A, F)$ be an algebra. An algebra of the form $(B, F|_B)$, where $B \subseteq A$ is invariant under all operations $f \in F$, is called a* subalgebra *of $\mathbb{A}$.*

As a subalgebra $(B, F|_B)$ of an algebra $(A, F)$ is determined by its universe $B$, we will at times use the universe $B$ to denote the subalgebra $(B, F|_B)$. We say that a subalgebra $(B, F|_B)$ of $(A, F)$ is a *proper* subalgebra if $B$ is a proper subset of $A$; and, is a *maximal proper* subalgebra if it is proper and the only subalgebra whose universe properly contains $B$ is $(A, F)$ itself.

Next, we give another construction of a new algebra from an existing algebra, namely, the construction of a *homomorphic image*.

**Definition 2.14** *Let $\mathbb{A} = (A, F)$ be an algebra. A* congruence *of $\mathbb{A}$ is an equivalence relation $\theta \subseteq A \times A$ that is invariant under all operations $f \in F$. When $\theta$ is a congruence of $\mathbb{A}$, the equivalence class of $\theta$ containing $a \in A$ is denoted $a^\theta$; and, for each operation $f \in F$, the operation $f^\theta$ given by $f^\theta(a_1^\theta, \ldots, a_k^\theta) = (f(a_1, \ldots, a_k))^\theta$, where $k$ denotes the arity of $f$, is well-defined. The set $A^\theta$ is defined as $\{a^\theta : a \in A\}$, and the set $F^\theta$ is defined as $\{f^\theta : f \in F\}$.*

**Definition 2.15** *Let $\mathbb{A} = (A, F)$ be an algebra. An algebra of the form $(A^\theta, F^\theta)$, where $\theta$ is a congruence of $\mathbb{A}$, is called a* homomorphic image *of $\mathbb{A}$.*

The following fact is known.

**Fact 2.16** *If $\mathbb{A}$ is an idempotent algebra and $\theta$ is a congruence of $\mathbb{A}$, then each equivalence class of $\theta$ is a subalgebra of $\mathbb{A}$.*

We give a proof of Fact 2.16 for completeness.

**Proof**. Let $f : A^k \to A$ be an operation of $\mathbb{A}$, let $B$ be an equivalence class of $\theta$, and let $b_1, \ldots, b_k$ be elements of $B$. We want to show that $f(b_1, \ldots, b_k) \in B$. Fix $b$ to be any element of $B$. We have $(b, b_1), \ldots, (b, b_k) \in \theta$. Since $\theta$ is a congruence of $\mathbb{A}$, the operation $f$ is a polymorphism of $\theta$, and $(f(b, \ldots, b), f(b_1, \ldots, b_k))$ is an element of $\theta$. By idempotence of $f$, we have $f(b, \ldots, b) = b$, from which it follows that $b$ and $f(b_1, \ldots, b_k)$ are in the same $\theta$-equivalence class and $f(b_1, \ldots, b_k) \in B$. □

We now combine the subalgebra and homomorphic image constructions to define the notion of a *factor*.

**Definition 2.17** *A* factor *of an algebra $\mathbb{A}$ is a homomorphic image of a subalgebra of $\mathbb{A}$.*

A known fact that we will make use of is that a factor of a factor of an algebra is a factor of the algebra.

## 2.3 Complexity

In this subsection, we discuss known results concerning the complexity of the problems $\mathsf{QCSP_c}(\Gamma)$ and $\mathsf{CSP_c}(\Gamma)$. We call a problem *tractable* if it is decidable in polynomial time, that is, it is in P, and the only notion of reduction we will use is many-one polynomial-time reduction.

The set of polymorphisms of a constraint language $\Gamma$ has been used to study the complexity of the problems $\mathsf{CSP_c}(\Gamma)$ and $\mathsf{QCSP_c}(\Gamma)$ [23, 21, 3, 7]. For instance, the following is known.

**Theorem 2.18** *(follows from [21, 3]) Let $\Gamma_1, \Gamma_2$ be finite constraint languages having the same idempotent polymorphisms. Then the problems $\mathsf{CSP_c}(\Gamma_1)$, $\mathsf{CSP_c}(\Gamma_2)$ reduce to each other, and likewise, the problems $\mathsf{QCSP_c}(\Gamma_1)$, $\mathsf{QCSP_c}(\Gamma_2)$ reduce to each other.*

Intuitively, Theorem 2.18 might be taken as saying that the polymorphisms of a constraint language $\Gamma$ contain all of the information needed to determine the complexity of $\Gamma$. In fact, it has been shown that when one considers the algebra having these polymorphisms as its operations, algebraic concepts such as those in the previous subsection can be employed to study complexity [7]. We will make use of this algebra and this approach.

**Definition 2.19** *When $\Gamma$ is a constraint language over set $A$, the algebra $\mathbb{A}_\Gamma$ is defined to be the algebra $(A, F)$ where $F$ is the set of idempotent polymorphisms of $\Gamma$.*

We will use the presence of idempotent polymorphisms to prove positive results concerning $\mathsf{QCSP_c}(\Gamma)$ complexity. In particular, we will make use of the following fact.

**Fact 2.20** *Let $\mathcal{C}$ be a finite conjunction of constraints over variables $\{v_1, \ldots, v_m\}$, assume that $f : A^k \to A$ is an idempotent polymorphism of all relations in $\mathcal{C}$, and let $g_1, \ldots, g_k : \{v_1, \ldots, v_m\} \to A$ be assignments satisfying $\mathcal{C}$. Then, the assignment $g : \{v_1, \ldots, v_m\} \to A$ defined by $g(v_i) = f(g_1(v_i), \ldots, g_k(v_i))$ for all $v_i$, satisfies $\mathcal{C}$.*

This fact is known; it is implicit, for instance, in [7]. We give a proof for completeness.

**Proof.** Let $R(w_1, \ldots, w_n)$ be a constraint where $f$ is a polymorphism of $R$ and that is satisfied by the assignments $g_1, \ldots, g_k$. It suffices to show that the assignment $g$ also satisfies $R$. For each $i \in [k]$, let $g'_i : A \cup \{v_1, \ldots, v_m\} \to A$ be the extension of $g_i$ that acts as the identity on $A$; likewise, let $g' : A \cup \{v_1, \ldots, v_m\} \to A$ be the extension of $g$ that acts as the identity on $A$. Since each $g_i$ satisfies the constraint, we have $(g'_i(w_1), \ldots, g'_i(w_n)) \in R$ for all $i \in [k]$. Since $f$ is a polymorphism of $R$, we have that $(f(g'_1(w_1), \ldots, g'_k(w_1)), \ldots, f(g'_1(w_n), \ldots, g'_k(w_n)))$ is contained in $R$. We claim that this tuple is equal to $(g'(w_1), \ldots, g'(w_n))$, which would give the proof. Suppose that $w_i$ is a variable. Then, we have $f(g'_1(w_i), \ldots, g'_k(w_i)) = f(g_1(w_i), \ldots, g_k(w_i)) = g(w_i) = g'(w_i)$. Suppose that $w_i$ is a constant. Then, we have $f(g'_1(w_i), \ldots, g'_k(w_i)) = f(w_i, \ldots, w_i) = w_i = g'(w_i)$, as $f$ is idempotent. □

To derive negative complexity results, we will make use of the following known result.

**Definition 2.21** *An algebra $(A, F)$ is a G-set if its universe is not one-element and every operation $f \in F$ is of the form $f(x_1, \ldots, x_k) = \pi(x_i)$ where $i \in [k]$ and $\pi$ is a permutation on $A$.*

We would like to emphasize that, in this paper, we require a G-set to be non-trivial.

**Theorem 2.22** *[7] Let $\Gamma$ be a constraint language. If $\mathbb{A}_\Gamma$ has a G-set as factor, then the problem $\mathsf{CSP_c}(\Gamma)$ is NP-complete, and hence the problem $\mathsf{QCSP_c}(\Gamma)$ is NP-hard.*

Theorem 2.22 shows that the absence of a G-set as a factor of $\mathbb{A}_\Gamma$ is necessary for the problem $\mathsf{CSP}_\mathsf{c}(\Gamma)$ to be tractable. It has been conjectured that this absence is also sufficient for $\mathsf{CSP}_\mathsf{c}(\Gamma)$ tractability [7]. The following classification results of Bulatov confirm this conjecture for two broad classes of constraint languages.

**Theorem 2.23** *(Bulatov [9]) Let $\Gamma$ be a constraint language on a three-element domain. If $\mathbb{A}_\Gamma$ does not contain a G-set as factor, then $\mathsf{CSP}_\mathsf{c}(\Gamma)$ is in P.*

**Theorem 2.24** *(Bulatov [5]) Let $\Gamma$ be a constraint language on a finite domain A containing all subsets of A. If $\mathbb{A}_\Gamma$ does not contain a G-set as factor, then $\mathsf{CSP}_\mathsf{c}(\Gamma)$ is in P.*

In the case of a two-element algebra, there is a nice description of idempotent algebras stating that either an algebra must contain one of four particular operations, or be a G-set. This description can be obtained from a classification theorem due to Post; see [2] for a presentation of this theorem. To give the description, we require a couple of definitions. When $A$ is a two-element set, we define the *majority operation on $A$*, denoted by $\mathsf{majority}_A : A^3 \to A$, to be the ternary operation satisfying the identities $\mathsf{majority}_A(x, x, y) = \mathsf{majority}_A(x, y, x) = \mathsf{majority}_A(y, x, x) = x$. That is, the operation $\mathsf{majority}_A$ returns the element of $A$ that occurs two or three times. Also, we define the *minority operation on $A$*, denoted by $\mathsf{minority}_A : A^3 \to A$, to be the ternary operation satisfying the identities $\mathsf{minority}_A(x, x, y) = \mathsf{minority}_A(x, y, x) = \mathsf{minority}_A(y, x, x) = y$. That is, the operation $\mathsf{minority}_A$ is idempotent, and if both elements of $A$ occur as arguments, it returns the element that occurs once.

**Theorem 2.25** *(see [2]) Let $\mathbb{A}$ be an idempotent algebra with universe $\{0, 1\}$. Either $\mathbb{A}$ is a G-set, or contains as term operation one of the following four operations:*

- *the binary AND operation $\wedge$,*
- *the binary OR operation $\vee$,*
- *the operation $\mathsf{majority}_{\{0,1\}}$,*
- *the operation $\mathsf{minority}_{\{0,1\}}$.*

We can now readily state the classification theorem for problems $\mathsf{QCSP}_\mathsf{c}(\Gamma)$ over a two-element domain.

**Theorem 2.26** *(follows from [12] and Theorem 2.25) Let $\Gamma$ be a constraint language on the two-element domain $\{0, 1\}$. If $\Gamma$ has as polymorphism one of the four operations given in the statement of Theorem 2.25, then $\mathsf{QCSP}_\mathsf{c}(\Gamma)$ is in P. Otherwise, $\mathbb{A}_\Gamma$ is a G-set and $\mathsf{QCSP}_\mathsf{c}(\Gamma)$ is PSPACE-complete.*

## 3 Collapsings

In this section, we introduce the definitions and ideas that lie at the base of our methodology for proving QCSP complexity results. Our methodology allows one to demonstrate that for certain constraint languages $\Gamma$, the problem $\mathsf{QCSP}_\mathsf{c}(\Gamma)$ can be reduced to the problem $\mathsf{CSP}_\mathsf{c}(\Gamma)$. More specifically, we will show that for certain constraint languages $\Gamma$, an instance of $\mathsf{QCSP}_\mathsf{c}(\Gamma)$ is true if and only if all instances in an ensemble of simpler instances of $\mathsf{QCSP}_\mathsf{c}(\Gamma)$ are true. The simpler instances are derived from the original instance by instantiating all but a bounded number of the universally quantified variables with a constant; we will show that these simpler instances can be formulated as an instance of $\mathsf{CSP}_\mathsf{c}(\Gamma)$. Their precise definition is as follows.

**Definition 3.1** *Let $\Phi$ be a quantified constraint formula with domain $A$. A quantified constraint formula $\Phi'$ is a $(j, a)$-collapsing of $\Phi$ if it can be obtained from $\Phi$ by choosing a subset $U'$ of the universally quantified variables $U^\Phi$ with size $|U'| \leq j$ and instantiating the variables in $U^\Phi \setminus U'$ with the constant $a \in A$. A quantified constraint formula $\Phi'$ is a $j$-collapsing of $\Phi$ if for some $a \in A$, the formula $\Phi'$ is a $(j, a)$-collapsing of $\Phi$. (Throughout this section, we assume $j \geq 0$.)*

By instantiating a variable $v$ of a quantified constraint formula with a constant, we mean that the variable and its quantifier are removed from the quantifier prefix, and that all instances of the variable in constraints are replaced with the constant.

**Example 3.2** *Consider the quantified constraint formula*

$$\Phi = \forall y_1 \exists x_1 \forall y_2 \forall y_3 \exists x_2 (R_1(y_1, x_1) \wedge R_2(y_2, x_2) \wedge R_3(y_2, x_2, y_3)).$$

*Suppose that its domain is $A = \{a, b\}$. There are four $(1, b)$-collapsings of $\Phi$, corresponding to the choices $U' = \{y_1\}$, $U' = \{y_2\}$, $U' = \{y_3\}$, and $U' = \emptyset$, respectively.*

$$\forall y_1 \exists x_1 \exists x_2 (R_1(y_1, x_1) \wedge R_2(b, x_2) \wedge R_3(b, x_2, b))$$
$$\exists x_1 \forall y_2 \exists x_2 (R_1(b, x_1) \wedge R_2(y_2, x_2) \wedge R_3(y_2, x_2, b))$$
$$\exists x_1 \forall y_3 \exists x_2 (R_1(b, x_1) \wedge R_2(b, x_2) \wedge R_3(b, x_2, y_3))$$
$$\exists x_1 \exists x_2 (R_1(b, x_1) \wedge R_2(b, x_2) \wedge R_3(b, x_2, b))$$

**Example 3.3** *An example of the type of result we will prove is the following. Suppose $\Gamma$ is a constraint language over the domain $\{0, 1\}$ having the boolean AND function $\wedge$ as polymorphism. Then, an instance $\Phi$ of $\mathsf{QCSP_c}(\Gamma)$ is true if and only if all $(1, 1)$-collapsings of $\Phi$ are true; this is proved below in Theorem 4.3.*

As discussed in Section 2.1, each quantified constraint formula can be viewed as a two-player game. Recall that in this game view, a *universal player* sets the universally quantified variables, and an *existential player* sets the existentially quantified variables. The existential player attempts to satisfy the constraints of the formula, while the universal player attempts to falsify a constraint. In order for us to prove results on the $(j, a)$-collapsings of a formula, it will be useful for us to consider a modified version of this game where the universal player has less power: each universally quantified variable $y$ has a subset of $A$ associated with it, and the universal player must set each universally quantified variable to a value falling within the subset associated to $y$. To formalize the idea of associating subsets of $A$ with universally quantified variables, we define the notion of *adversary*.

**Definition 3.4** *An* adversary *$\mathcal{A}$ of length $n$ is a tuple $\mathcal{A} \in (\wp(A) \setminus \{\emptyset\})^n$. We use $\mathcal{A}_i$ to denote the $i$th coordinate of the adversary $\mathcal{A}$, that is, $\mathcal{A} = (\mathcal{A}_1, \ldots, \mathcal{A}_n)$.*

Let us say that an adversary $\mathcal{A}$ is an adversary *for* a quantified constraint formula $\Phi$ if the length of $\mathcal{A}$ matches the number of universally quantified variables in $\Phi$. When this is the case, the adversary $\mathcal{A}$ naturally induces the set of assignments $\mathcal{A}[\Phi] = \{\tau : U^\Phi \to A : \tau(y_i) \in \mathcal{A}_i \text{ for all } i \in [n]\}$. Here, we assume that $y_1, \ldots, y_n$ are the universally quantified variables of $\Phi$, ordered according to quantifier prefix, from outside to inside.

We say that an adversary is $\Phi$-*winnable* if in the modified game, the existential player can win: that is, if there is a strategy that can handle all assignments that the adversary gives rise to, as formalized in the following definition.

**Definition 3.5** *Let $\Phi$ be a quantified constraint formula, and let $\mathcal{A}$ be an adversary for $\Phi$. We say that $\mathcal{A}$ is $\Phi$-winnable if there exists a strategy $\sigma$ for $\Phi$ such that for all assignments $\tau \in \mathcal{A}[\Phi]$, the assignment $\langle \sigma, \tau \rangle$ is defined and satisfies the constraints of $\Phi$.*

We have previously given a characterization of truth for quantified constraint formulas (Fact 2.6). This characterization can be formulated in the terminology just introduced. Let $A^n$ denote the adversary $(A, \ldots, A)$ of length $n$ that is equal to $A$ at all coordinates.

**Fact 3.6** *The adversary $A^n$ is $\Phi$-winnable if and only if $\Phi$ is true.*

**Proof.** Immediate from Fact 2.6 and Definition 3.5. □

In general, when $B$ is a non-empty subset of $A$, we will use $B^n$ to denote the adversary $(B, \ldots, B)$ of length $n$ that is equal to $B$ at all coordinates.

Fact 3.6 shows that the truth of a quantified constraint formula can be characterized in terms of an adversary. The truth of the $j$-collapsings of a quantified constraint formula also have an adversary-based characterization. To give this characterization, we introduce the following notation, which will be useful throughout the paper.

**Definition 3.7** *Let $n \geq 1$, $S \subseteq [n]$, and $C, D \subseteq A$. We use $\mathcal{A}(n, C, S, D)$ to denote the length $n$ adversary equal to $D$ at the coordinates in $S$, and equal to $C$ at all other coordinates. That is,*

- $\mathcal{A}(n, C, S, D)_i = C$ *for $i \in [n] \setminus S$, and*
- $\mathcal{A}(n, C, S, D)_i = D$ *for $i \in S$.*

*For $w \geq 1$, we define $\mathsf{Adv}(n, C, w, D) = \{\mathcal{A}(n, C, S, D) : S \subseteq [n], |S| \leq w\}$. That is, $\mathsf{Adv}(n, C, w, D)$ is the set of all length $n$ adversaries that are equal to $D$ in at most $w$ coordinates, and equal to $C$ at all other coordinates.*

Note that when using the $\mathsf{Adv}(n, C, w, D)$ notation, we will typically have $w \leq n$ and $C \subseteq D$.

**Example 3.8** *Suppose that $A = \{0, 1\}$. The set $\mathsf{Adv}(4, \{1\}, 1, A)$ is equal to*

$$\begin{aligned}\{\ & (A, \{1\}, \{1\}, \{1\}), \\ & (\{1\}, A, \{1\}, \{1\}), \\ & (\{1\}, \{1\}, A, \{1\}), \\ & (\{1\}, \{1\}, \{1\}, A), \\ & (\{1\}, \{1\}, \{1\}, \{1\})\ \}\end{aligned}$$

The following is our adversary-based characterization of the $j$-collapsings of a quantified constraint formula.

**Proposition 3.9** *Let $\Phi$ be a quantified constraint formula. The $(j, a)$-collapsings of $\Phi$ are true if and only if the adversaries $\mathsf{Adv}(n, \{a\}, j, A)$ are $\Phi$-winnable. And, the $j$-collapsings of $\Phi$ are true if and only if the adversaries $\cup_{a \in A} \mathsf{Adv}(n, \{a\}, j, A)$ are $\Phi$-winnable.*

Proposition 3.9 can be proved in a straightforward fashion by using Fact 3.6. We omit the proof.

**Remark 3.10** *Let us say that an adversary $\mathcal{A}$ of length $n$ is* dominated by *a second adversary $\mathcal{B}$ of length $n$ if $\mathcal{A}_i \subseteq \mathcal{B}_i$ for all $i \in [n]$. For example, the adversary $(\{1\}, \{1\}, \{1\}, \{1\})$ in Example 3.8 is dominated by all other adversaries in that example. It is readily seen that if $\mathcal{A}$ is dominated by $\mathcal{B}$, then $\mathcal{A}$ is easier than $\mathcal{B}$ in the sense that the $\Phi$-winnability of $\mathcal{B}$ implies the $\Phi$-winnability of $\mathcal{A}$. Thus, the adversary $(\{1\}, \{1\}, \{1\}, \{1\})$ in Example 3.8 is in a sense superfluous as a member of $\mathsf{Adv}(4, \{1\}, 1, A)$: omitting it from this set of adversaries would not change the class of quantified constraint formulas $\Phi$ such that $\mathsf{Adv}(4, \{1\}, 1, A)$ is*

Φ-winnable. In general, if we defined $\mathsf{Adv}(n, C, w, D)$ as $\{\mathcal{A}(n, C, S, D) : S \subseteq [n], |S| = w\}$, demanding exactly $w$ coordinates of each adversary to be equal to $D$, then Proposition 3.9 would still hold (assuming $j \leq n$). We elected the given definition of $\mathsf{Adv}(n, C, w, D)$ because it will facilitate the presentation of certain proofs.

To review, our goal is to show that for certain constraint languages $\Gamma$, the problem $\mathsf{QCSP}_\mathsf{c}(\Gamma)$ is reducible to $\mathsf{CSP}_\mathsf{c}(\Gamma)$. We will achieve this, for a constraint language, by showing that the constraint language satisfies a property that we call *collapsibility*.

**Definition 3.11** *A constraint language $\Gamma$ is $(j, a)$-collapsible when the following holds: if all $(j, a)$-collapsings of an instance $\Phi$ of $\mathsf{QCSP}_\mathsf{c}(\Gamma)$ are true, then the instance $\Phi$ is true. Similarly, a constraint language $\Gamma$ is $j$-collapsible when the following holds: if all $j$-collapsings of an instance $\Phi$ of $\mathsf{QCSP}_\mathsf{c}(\Gamma)$ are true, then the instance $\Phi$ is true.*

When a quantified constraint formula $\Phi$ is true, then all $j$-collapsings of $\Phi$ are true; one way to see this is that all adversaries, and hence in particular those identified in Proposition 3.9, are dominated (in the sense of Remark 3.10) by the adversary $A^n$. Thus, if a constraint language $\Gamma$ is $j$-collapsible, an instance of $\mathsf{QCSP}_\mathsf{c}(\Gamma)$ is true *if and only if* all of its $j$-collapsings are true. The following proposition shows that, starting from an instance $\Phi$ of $\mathsf{QCSP}_\mathsf{c}(\Gamma)$, the truth of the $j$-collapsings of $\Phi$ can be efficiently translated into an instance of $\mathsf{CSP}_\mathsf{c}(\Gamma)$, and so the property of $j$-collapsibility implies a reduction from $\mathsf{QCSP}_\mathsf{c}(\Gamma)$ to $\mathsf{CSP}_\mathsf{c}(\Gamma)$.

**Proposition 3.12** *If there exists $j \geq 0$ such that the constraint language $\Gamma$ is $j$-collapsible (or, there exists $j \geq 0$ and $a \in A$ such that the constraint language is $(j, a)$-collapsible), then the problem $\mathsf{QCSP}_\mathsf{c}(\Gamma)$ reduces to (and is hence equivalent to) $\mathsf{CSP}_\mathsf{c}(\Gamma)$.*

In order to prove Proposition 3.12, we will make use of the following lemma.

**Lemma 3.13** *For each fixed $j \geq 0$ and constraint language $\Gamma$, there is a polynomial-time algorithm that computes, given an instance $\Phi$ of $\mathsf{QCSP}_\mathsf{c}(\Gamma)$ and a $j$-collapsing $\Phi'$ of $\Phi$, an instance $\Phi''$ of $\mathsf{CSP}_\mathsf{c}(\Gamma)$ that is true if and only if $\Phi'$ is true.*

**Proof**. The idea of the proof is to create an instance $\Phi''$ of $\mathsf{CSP}_\mathsf{c}(\Gamma)$ where the variables are the possible output values of a strategy for $\Phi'$. The instance $\Phi''$ will be true if and only if there is a winning strategy for $\Phi'$.

The variables of $\Phi''$ are all pairs of the form $(x, \alpha)$, where $x \in E^{\Phi'}$ is an existentially quantified variable of $\Phi'$ and $\alpha$ is a mapping from $U_x^{\Phi'}$ to $A$. For every constraint $R(v_1, \ldots, v_m)$ and every mapping $\tau : U^{\Phi'} \to A$, we place the constraint $R(w_1, \ldots, w_m)$ in $\Phi''$, where for each $i \in [m]$ we define $w_i = \tau(v_i)$ if $v_i$ is universally quantified in $\Phi'$, and $w_i = (v_i, \tau|_{U_{v_i}^{\Phi'}})$ if $v_i$ is existentially quantified in $\Phi'$.

Suppose that $f : V^{\Phi''} \to A$ is an assignment to the variables of $\Phi''$. It is straightforward to verify that $f$ satisfies the constraints of $\Phi''$ if and only if the strategy $\sigma$ defined by $\sigma_x(\alpha) = f((x, \alpha))$ for all $x \in E^{\Phi'}$ and mappings $\alpha : U_x^{\Phi'} \to A$, is a winning strategy.

As $\Phi'$ has $j$ or fewer universally quantified variables and $A$ is fixed, the number of mappings from $U^{\Phi'}$ to $A$ (and from $U_x^{\Phi'}$ to $A$ for any $x \in E^{\Phi'}$) is bounded above by a constant. Having observed this, it is clear that $\Phi''$ can be computed from $\Phi'$ in polynomial time. $\square$

**Proof**. (Proposition 3.12) We prove the proposition for a $j$-collapsible constraint language $\Gamma$; the proof is similar for a $(j, a)$-collapsible constraint language.

Let $\Phi$ be an instance of $\mathsf{QCSP}_\mathsf{c}(\Gamma)$. As discussed prior to the statement of this proposition, the formula $\Phi$ is true if and only if all $j$-collapsings of $\Phi$ are true. We can compute all $j$-collapsings of $\Phi$ in polynomial time. By Lemma 3.13, each of these collapsings can be converted to an instance of $\mathsf{CSP}_c(\Gamma)$ in polynomial time. It remains to show that all of the resulting $\mathsf{CSP}_c(\Gamma)$ instances $\{\Phi_1, \ldots, \Phi_m\}$ can be formulated as a single $\mathsf{CSP}_c(\Gamma)$ instance. This can be done by renaming variables so that no two instances among $\{\Phi_1, \ldots, \Phi_m\}$ have a variable with the same name, and then creating a single instance whose constraints are all of the constraints appearing in one of the instances $\Phi_i$. $\square$

It is readily verified from the definitions in this section that if a constraint language is $j$-collapsible, then it is $j'$-collapsible for all $j' > j$. In addition, it is clear that the reduction from $\mathsf{QCSP}_c(\Gamma)$ to $\mathsf{CSP}_c(\Gamma)$ given by the proofs of Proposition 3.12 and Lemma 3.13 are more efficient for lower values of $j$ than for higher values of $j$. Thus, $j$-collapsibility results for lower values of $j$ are preferable to such results for higher values of $j$, although establishing the $j$-collapsibility of a constraint language $\Gamma$ for *any* value of $j$ implies that there is a reduction from $\mathsf{QCSP}_c(\Gamma)$ to $\mathsf{CSP}_c(\Gamma)$ (Proposition 3.12).

## 4  Composing Adversaries

In the previous section, we identified the properties of $j$-collapsibility and $(j,a)$-collapsibility; we showed that when a constraint language $\Gamma$ has one of these properties, $\mathsf{QCSP}_c(\Gamma)$ can be reduced to $\mathsf{CSP}_c(\Gamma)$. Our goal now is to develop and deploy machinery that allows one to prove collapsibility results–results showing that a constraint language is either $j$- or $(j,a)$-collapsible. To prove a collapsibility result, by definition (see Definition 3.11), we need to prove the truth of a formula based on the truth of its collapsings. We will accomplish this in the language of adversaries. In particular, we know that the truth of collapsings can be characterized by the winnability of certain adversaries (Proposition 3.9), while the truth of a quantified constraint formula can be characterized by the winnability of the "full adversary" $A^n$ (Fact 3.6). Thinking of the adversaries corresponding to collapsings as simple adversaries, our method will be to assume the winnability of these simple adversaries, and then derive the winnability of more and more complex adversaries, until the winnability of the full adversary is derived.

In this section, we present a notion of composition that will allow us to derive the winnability of more complex adversaries from simpler ones. After this, we give examples of how this notion of composition can be used to prove collapsibility results, and hence derive positive $\mathsf{QCSP}_c(\Gamma)$ complexity results. We begin by describing our notion of composition.

Let $f : A^k \to A$ be an operation and let $\mathcal{A}, \mathcal{B}_1, \ldots, \mathcal{B}_k$ be adversaries of length $n$. We say that $\mathcal{A}$ is $f$-*composable* from $\mathcal{B}_1, \ldots, \mathcal{B}_k$, denoted $\mathcal{A} \triangleleft f(\mathcal{B}_1, \ldots, \mathcal{B}_k)$, if for all $i \in [n]$, it holds that $\mathcal{A}_i \subseteq f(\mathcal{B}_{1i}, \ldots, \mathcal{B}_{ki})$. The following key theorem allows us to derive the winnability of an adversary based on the winnability of adversaries that it can be composed from.

**Theorem 4.1** *Let $\Phi$ be a quantified constraint formula, assume that $f : A^k \to A$ is an idempotent polymorphism of all relations of $\Phi$, and let $\mathcal{A}, \mathcal{B}_1, \ldots, \mathcal{B}_k$ be adversaries for $\Phi$. If each of the adversaries $\mathcal{B}_1, \ldots, \mathcal{B}_k$ is $\Phi$-winnable and $\mathcal{A} \triangleleft f(\mathcal{B}_1, \ldots, \mathcal{B}_k)$, then the adversary $\mathcal{A}$ is $\Phi$-winnable.*

**Proof**. For all $j \in [k]$, let $\sigma^j$ be a strategy witnessing the $\Phi$-winnability of $\mathcal{B}_j$. Since $\mathcal{A} \triangleleft f(\mathcal{B}_1, \ldots, \mathcal{B}_k)$, there exist functions $g_i^j : \mathcal{A}_i \to \mathcal{B}_{ji}$ (with $i \in [n]$, $j \in [k]$) such that for all $i \in [n]$ and $a \in \mathcal{A}_i$, it holds that $a = f(g_i^1(a), \ldots, g_i^k(a))$. Let us denote the universal variables of $\Phi$ by $y_1, \ldots, y_n$, and assume that they are ordered according to the quantifier prefix, as in the discussion before Definition 3.5.

The idea of the proof is this. We would like to construct a strategy $\sigma$ for the adversary $\mathcal{A}$ based on the strategies $\sigma^j$ for the adversaries $\mathcal{B}_j$. The strategy $\sigma$ simulates the strategies $\sigma^j$. Upon being given a value $a \in \mathcal{A}_i$ for the universally quantified variable $y_i$, the strategy $\sigma$ passes the values $g_i^1(a), \ldots, g_i^k(a)$, which form an "inverse" of $a$ under $f$, to the strategies $\sigma^1, \ldots, \sigma^k$. When the strategy $\sigma$ needs to set an existentially quantified variable $x$, it takes the assignments given to $x$ by $\sigma^1, \ldots, \sigma^k$ and applies $f$ to them to obtain its setting. The assignment produced by $\sigma$ will be equal to the assignments produced by $\sigma^1, \ldots, \sigma^k$ mapped under $f$ (applied point-wise). We now make this idea precise.

When $\tau$ is a function in $\mathcal{A}[\Phi]$, or the restriction of a function in $\mathcal{A}[\Phi]$, for all $j \in [k]$, we define $g^j(\tau)$ to be the function with the same domain as $\tau$ and where $g^j(\tau)(y_i) = g_i^j(\tau(y_i))$ for all elements $y_i$ of this domain. Observe that for all $\tau \in \mathcal{A}[\Phi]$ and $j \in [k]$, it holds that $g^j(\tau) \in \mathcal{B}_j[\Phi]$.

We define the strategy $\sigma$ for $\mathcal{A}$ as follows. For all $x \in E^\Phi$, define $\sigma_x$ by

$$\sigma_x(\tau) = f(\sigma_x^1(g^1(\tau)), \ldots, \sigma_x^k(g^k(\tau)))$$

for all functions $\tau : U_x^\Phi \to A$ that arise as the restriction of a function in $\mathcal{A}[\Phi]$ to $U_x^\Phi$.

We claim that for all $\tau \in \mathcal{A}[\Phi]$ and for all variables $v \in V^\Phi$,

$$\langle \sigma, \tau \rangle(v) = f(\langle \sigma^1, g^1(\tau) \rangle(v), \ldots, \langle \sigma^k, g^k(\tau) \rangle(v)).$$

Showing this claim suffices to give the theorem by Fact 2.20 and the assumption that the $\sigma^j$ are winning strategies for the $\mathcal{B}_j$.

For universal variables $y_i \in U^\Phi$, we have

$$\begin{aligned}
\langle \sigma, \tau \rangle(y_i) &= \tau(y_i) \\
&= f(g_i^1(\tau(y_i)), \ldots, g_i^k(\tau(y_i))) \\
&= f(g^1(\tau)(y_i), \ldots, g^k(\tau)(y_i)) \\
&= f(\langle \sigma^1, g^1(\tau) \rangle(y_i), \ldots, \langle \sigma^k, g^k(\tau) \rangle(y_i))
\end{aligned}$$

For existential variables $x \in E^\Phi$, we have

$$\begin{aligned}
\langle \sigma, \tau \rangle(x) &= \sigma_x(\tau|_{U_x^\Phi}) \\
&= f(\sigma_x^1(g^1(\tau|_{U_x^\Phi})), \ldots, \sigma_x^k(g^k(\tau|_{U_x^\Phi}))) \\
&= f(\sigma_x^1(g^1(\tau)|_{U_x^\Phi}), \ldots, \sigma_x^k(g^k(\tau)|_{U_x^\Phi})) \\
&= f(\langle \sigma^1, g^1(\tau) \rangle(x), \ldots, \langle \sigma^k, g^k(\tau) \rangle(x))
\end{aligned}$$

□

**Remark 4.2** *The hypothesis of Theorem 4.1 that $f$ is an* idempotent *polymorphism can be removed if we are concerned with quantified constraint formulas where constants do* not *appear in relations. Indeed, the only place in the proof of Theorem 4.1 where the idempotence of $f$ is used is in the appeal to Fact 2.20, and this fact holds for non-idempotent polymorphisms when the constraints do not contain constants.*

Armed with Theorem 4.1, we now present three examples of collapsibility results. The first concerns the boolean AND function as a polymorphism.

**Theorem 4.3** *Let $\Gamma$ be a constraint language over the domain $A = \{0, 1\}$ having the boolean AND function $\wedge$ as polymorphism. The constraint language $\Gamma$ is $(1, 1)$-collapsible.*

**Proof**. Let $\Phi$ be an instance of $\mathsf{QCSP}_\mathsf{c}(\Gamma)$, and assume that the $(1, 1)$-collapsings of $\Phi$ are true. By Proposition 3.9, the adversaries $\mathsf{Adv}(n, \{1\}, 1, A)$ are $\Phi$-winnable. We want to prove that the instance $\Phi$ is true; by Fact 3.6, it suffices to prove that the adversary $A^n$ is $\Phi$-winnable.

We prove by induction that for all $i \in [n]$, the adversary $\mathcal{A}(n, \{1\}, [i], A)$ is $\Phi$-winnable.

For $i = 1$, this holds by hypothesis as $\mathcal{A}(n, \{1\}, [1], A) \in \text{Adv}(n, \{1\}, 1, A)$.

For the induction, let us assume that the adversary $\mathcal{A}(n, \{1\}, [i], A)$ is $\Phi$-winnable for a value $i < n$. Observe that the adversary $\mathcal{A}(n, \{1\}, \{i+1\}, A)) \in \text{Adv}(n, \{1\}, 1, A)$ is $\Phi$-winnable by hypothesis. We claim that
$$\mathcal{A}(n, \{1\}, [i+1], A) \triangleleft \wedge(\mathcal{A}(n, \{1\}, [i], A), \mathcal{A}(n, \{1\}, \{i+1\}, A))$$
which, by appeal to Theorem 4.1, gives the induction. We verify that
$$\mathcal{A}(n, \{1\}, [i+1], A)_j \subseteq \wedge(\mathcal{A}(n, \{1\}, [i], A)_j, \mathcal{A}(n, \{1\}, \{i+1\}, A)_j)$$
for all $j \in [n]$ as follows:

- For $j \in [i]$, we have $A \subseteq \wedge(A, \{1\})$, since $a = \wedge(a, 1)$ for all $a \in A$.

- For $j = i+1$, we have $A \subseteq \wedge(\{1\}, A)$, since $a = \wedge(1, a)$ for all $a \in A$.

- For $j \in [n] \setminus [i+1]$, we have $\{1\} \subseteq \wedge(\{1\}, \{1\})$.

□

The next two examples each have a similar proof structure. We assume the winnability of the "simple" adversaries corresponding to the relevant collapsings, and derive the winnability of more and more complex adversaries in an inductive manner, ultimately deriving the winnability of the "full" adversary $A^n$.

**Definition 4.4** *A* Mal'tsev operation *is a ternary operation* $m : A^3 \to A$ *satisfying the identities* $m(x, x, y) = m(y, x, x) = y$.

**Example 4.5** *Let $A$ be a two-element set. The operation* $\text{minority}_A$*, defined in Section 2.3, is an example of a Mal'tsev operation.*

**Theorem 4.6** *Let $\Gamma$ be a constraint language having a Mal'tsev operation $m : A^3 \to A$ as polymorphism. The constraint language $\Gamma$ is $(1, a)$-collapsible for any $a \in A$.*

**Proof**. Let $\Phi$ be an instance of $\text{QCSP}_c(\Gamma)$, fix $a \in A$, and assume that the $(1, a)$-collapsings of $\Phi$ are true. By Proposition 3.9, we have that the adversaries $\text{Adv}(n, \{a\}, 1, A)$ are $\Phi$-winnable, and by Fact 3.6, it suffices to prove that the adversary $A^n$ is $\Phi$-winnable.

We prove by induction that for all $i \in [n]$, the adversary $\mathcal{A}(n, \{a\}, [i], A)$ is $\Phi$-winnable.

For $i = 1$, this holds by hypothesis as $\mathcal{A}(n, \{a\}, [1], A) \in \text{Adv}(n, \{a\}, 1, A)$.

For the induction, let us assume that the adversary $\mathcal{A}(n, \{a\}, [i], A)$ is $\Phi$-winnable for a value $i < n$. Observe that the adversary $\mathcal{A}(n, \{a\}, \{i+1\}, A)) \in \text{Adv}(n, \{a\}, 1, A)$ is $\Phi$-winnable by hypothesis. We claim that
$$\mathcal{A}(n, \{a\}, [i+1], A) \triangleleft m(\mathcal{A}(n, \{a\}, [i], A), \mathcal{A}(n, \{a\}, \emptyset, A), \mathcal{A}(n, \{a\}, \{i+1\}, A))$$
which, by appeal to Theorem 4.1, gives the induction. We verify that
$$\mathcal{A}(n, \{a\}, [i+1], A)_j \subseteq m(\mathcal{A}(n, \{a\}, [i], A)_j, \mathcal{A}(n, \{a\}, \emptyset, A)_j, \mathcal{A}(n, \{a\}, \{i+1\}, A)_j)$$
for all $j \in [n]$ as follows:

- For $j \in [i]$, we have $A \subseteq m(A, \{a\}, \{a\})$, since $b = m(b, a, a)$ for all $b \in A$.

- For $j = i+1$, we have $A \subseteq m(\{a\}, \{a\}, A)$, since $b = m(a, a, b)$ for all $b \in A$.

- For $j \in [n] \setminus [i+1]$, we have $\{a\} \subseteq m(\{a\}, \{a\}, \{a\})$.

□

**Definition 4.7** *The* dual discriminator operation *on a set $A$ is the ternary operation $d : A^3 \to A$ defined by*

$$d(x, y, z) = \begin{cases} x & \text{if } x = y \\ z & \text{otherwise.} \end{cases}$$

**Example 4.8** *Let $A$ be a two-element set. The operation* majority$_A$, *defined in Section 2.3, is the dual discriminator operation on $A$.*

**Theorem 4.9** *Let $\Gamma$ be a constraint language having the dual discriminator operation $d : A^3 \to A$ as polymorphism. The constraint language $\Gamma$ is 1-collapsible.*

**Proof.** Let $\Phi$ be an instance of $\mathsf{QCSP}_\mathsf{c}(\Gamma)$, and assume that the 1-collapsings of $\Phi$ are true. By Proposition 3.9, we have that the adversaries $\cup_{a \in A} \mathsf{Adv}(n, \{a\}, 1, A)$ are $\Phi$-winnable, and by Fact 3.6, it suffices to prove that the adversary $A^n$ is $\Phi$-winnable.

Fix two distinct elements $b, c \in A$. We prove by induction that for all $i \in [n]$, the adversaries $\mathcal{A}(n, \{b\}, [i], A)$ and $\mathcal{A}(n, \{c\}, [i], A)$ are $\Phi$-winnable.

For $i = 1$, this holds by hypothesis as $\mathcal{A}(n, \{b\}, [1], A), \mathcal{A}(n, \{c\}, [1], A) \in \cup_{a \in A}\mathsf{Adv}(n, \{a\}, 1, A)$.

For the induction, let us assume that the adversaries $\mathcal{A}(n, \{b\}, [i], A)$ and $\mathcal{A}(n, \{c\}, [i], A)$ are $\Phi$-winnable for a value $i < n$. Observe that the adversaries

$$\mathcal{A}(n, \{b\}, \{i+1\}, A)), \mathcal{A}(n, \{c\}, \{i+1\}, A)) \in \cup_{a \in A}\mathsf{Adv}(n, \{a\}, 1, A)$$

are $\Phi$-winnable by hypothesis. We claim that

$$\mathcal{A}(n, \{b\}, [i+1], A) \triangleleft d(\mathcal{A}(n, \{b\}, [i], A), \mathcal{A}(n, \{c\}, [i], A), \mathcal{A}(n, \{b\}, \{i+1\}, A))$$

and

$$\mathcal{A}(n, \{c\}, [i+1], A) \triangleleft d(\mathcal{A}(n, \{c\}, [i], A), \mathcal{A}(n, \{b\}, [i], A), \mathcal{A}(n, \{c\}, \{i+1\}, A))$$

which, by appeal to Theorem 4.1, gives the induction. We explicitly verify the first of the two claims; the second is identical, but with the roles of $b$ and $c$ swapped. We verify that

$$\mathcal{A}(n, \{b\}, [i+1], A)_j \subseteq d(\mathcal{A}(n, \{b\}, [i], A)_j, \mathcal{A}(n, \{c\}, [i], A)_j, \mathcal{A}(n, \{b\}, \{i+1\}, A)_j)$$

for all $j \in [n]$ as follows:

- For $j \in [i]$, we have $A \subseteq d(A, A, \{b\})$, since $a = d(a, a, b)$ for all $a \in A$.

- For $j = i+1$, we have $A \subseteq d(\{b\}, \{c\}, A)$, since $a = d(b, c, a)$ for all $a \in A$.

- For $j \in [n] \setminus [i+1]$, we have $\{b\} \subseteq d(\{b\}, \{c\}, \{b\})$.

□

The polymorphisms addressed by the preceding three theorems are all known to imply $\mathsf{CSP}_\mathsf{c}(\Gamma)$ tractability. For instance, the following theorem is known.

**Theorem 4.10** *[4, 15] Let $\Gamma$ be a constraint language having a Mal'tsev operation $m : A^3 \to A$ as polymorphism. The problem $\mathsf{CSP}_\mathsf{c}(\Gamma)$ is polynomial-time tractable.*

Similarly, the $\mathsf{CSP}_\mathsf{c}(\Gamma)$ tractability of the binary AND operation is implied by [23, Theorem 5.13], and the tractability of the dual discriminator operation is implied by [23, Theorem 5.7]. Using these tractability results in conjunction with our collapsibility theorems, we obtain $\mathsf{QCSP}_\mathsf{c}(\Gamma)$ tractability results. For instance, we have the following theorem.

**Theorem 4.11** *Let $\Gamma$ be a constraint language having a Mal'tsev operation $m : A^3 \to A$ as polymorphism. The problem $\mathsf{QCSP}_\mathsf{c}(\Gamma)$ is polynomial-time tractable.*

**Proof**. By Theorem 4.6 and Proposition 3.12, the problem $\mathsf{QCSP}_\mathsf{c}(\Gamma)$ reduces to the problem $\mathsf{CSP}_\mathsf{c}(\Gamma)$. By Theorem 4.10, the problem $\mathsf{CSP}_\mathsf{c}(\Gamma)$ is polynomial-time tractable. The theorem follows. $\square$

The collapsibility theorems that we have just given also have a consequence for the problems $\mathsf{QCSP}_\mathsf{c}(\Gamma)$ over a two-element domain. Namely, if such a problem $\mathsf{QCSP}_\mathsf{c}(\Gamma)$ is tractable at all, then it is 1-collapsible. The property of collapsibility thus exposes uniform structure among such tractable problems $\mathsf{QCSP}_\mathsf{c}(\Gamma)$.

**Corollary 4.12** *Let $\Gamma$ be a constraint language over a two-element domain. If $\mathsf{QCSP}_\mathsf{c}(\Gamma)$ is polynomial-time tractable, then $\Gamma$ is 1-collapsible (assuming that P does not equal PSPACE).*

**Proof**. Without loss of generality, we may assume that the domain of $\Gamma$ is $A = \{0, 1\}$. By Theorem 2.26, if $\mathsf{QCSP}_\mathsf{c}(\Gamma)$ is polynomial-time tractable, then $\Gamma$ has as polymorphism one of the operations $\{\wedge, \vee, \mathsf{majority}_{\{0,1\}}, \mathsf{minority}_{\{0,1\}}\}$. If $\Gamma$ has the operation $\wedge$ as polymorphism, it is 1-collapsible by Theorem 4.3. The operation $\vee$ is equivalent to $\wedge$ with the roles of 0 and 1 reversed, so if $\Gamma$ has the operation $\vee$ as polymorphism, it is 1-collapsible by an identical proof. If $\Gamma$ has the operation $\mathsf{majority}_{\{0,1\}}$ as polymorphism, it is 1-collapsible by Theorem 4.9; and, If $\Gamma$ has the operation $\mathsf{minority}_{\{0,1\}}$ as polymorphism, it is 1-collapsible by Theorem 4.6. $\square$

## 5 Collapsibility

In the previous section, we justified our definition of collapsibility for constraint languages by giving examples of collapsibility results. In Section 5.1, we translate this definition into the language of algebras. We give a definition of what it means for an *algebra* to be collapsible, and then demonstrate that the collapsibility results on constraint languages $\Gamma$ given in the previous section can in fact be interpreted as collapsibility results on their associated algebras $\mathbb{A}_\Gamma$. The advantage of having this algebraic formulation is that we are able to establish powerful and general tools for deriving collapsibility results. In each of Sections 5.2, 5.3, and 5.4, we present a technique for deriving algebraic collapsibility results, and illustrate its use.

Throughout this section, $\mathbb{A} = (A, F)$ denotes a finite idempotent algebra.

### 5.1 Collapsibility for Algebras

Let us first define what it means for an algebra to be collapsible. Say that an adversary $\mathcal{A}$ is $\mathbb{A}$-*composable* from a set of adversaries $\mathcal{S}$ if there exists a term operation $f : A^k \to A$ of $\mathbb{A}$ and adversaries $\mathcal{B}_1, \ldots, \mathcal{B}_k \in \mathcal{S}$ such that $\mathcal{A}$ is $f$-composable from $\mathcal{B}_1, \ldots, \mathcal{B}_k$, that is, $\mathcal{A} \triangleleft f(\mathcal{B}_1, \ldots, \mathcal{B}_k)$.

**Definition 5.1** *An algebra $\mathbb{A} = (A, F)$ is* collapsible *with source $S \subseteq A$ and width $w \geq 0$ if for all $n \geq 1$, the adversary $A^n$ is $\mathbb{A}$-composable from the set of adversaries $\cup_{a \in S} \mathsf{Adv}(n, \{a\}, w, A)$.*

When an algebra is collapsible, we will sometimes not state the source or width. For instance, we will say that an algebra $\mathbb{A}$ is collapsible with source $S \subseteq A$ if there exists a width $w \geq 0$ such that $\mathbb{A}$ is collapsible with source $S$ and width $w$.

We now relate this definition of collapsibility for algebras to the definition of collapsibility for constraint languages, showing that the collapsibility of the algebra $\mathbb{A}_\Gamma$ associated to a constraint language $\Gamma$ implies the collapsibility of $\Gamma$ itself.

**Proposition 5.2** *Let $\Gamma$ be a constraint language. If $\mathbb{A}_\Gamma$ is collapsible with width $w \geq 0$, then $\Gamma$ is $w$-collapsible, and hence the problem $\mathsf{QCSP}_\mathsf{c}(\Gamma)$ reduces to $\mathsf{CSP}_\mathsf{c}(\Gamma)$. Also, if $\mathbb{A}_\Gamma$ is collapsible with width $w \geq 0$ and source $\{a\}$, for some $a \in A$, then $\Gamma$ is $(w,a)$-collapsible, and hence the problem $\mathsf{QCSP}_\mathsf{c}(\Gamma)$ reduces to $\mathsf{CSP}_\mathsf{c}(\Gamma)$.*

**Proof**. We prove the first part; proof of the second part is similar. Suppose that all $j$-collapsings of an instance $\Phi$ of $\mathsf{QCSP}_\mathsf{c}(\Gamma)$ are true. We want to show that the instance $\Phi$ is true. By Proposition 3.9, the adversaries $\cup_{a \in A}\mathsf{Adv}(n, \{a\}, j, A)$ are $\Phi$-winnable. By hypothesis, the adversary $A^n$ is $\mathbb{A}$-composable from the adversaries $\cup_{a \in A}\mathsf{Adv}(n, \{a\}, j, A)$. Thus, by Theorem 4.1, the adversary $A^n$ is $\Phi$-winnable. By Fact 3.6, the formula $\Phi$ is true. $\square$

We now show that the notion of $\mathbb{A}$-composability, on which the definition of collapsibility for algebras is based, is robust in that it satisfies a certain type of transitivity.

**Proposition 5.3** *Let $\mathcal{S}$ and $\mathcal{S}'$ be sets of adversaries, all of the same length. If an adversary $\mathcal{A}$ is $\mathbb{A}$-composable from $\mathcal{S}'$, and all adversaries in $\mathcal{S}'$ are $\mathbb{A}$-composable from $\mathcal{S}$, then $\mathcal{A}$ is $\mathbb{A}$-composable from $\mathcal{S}$.*

**Proof**. By hypothesis, we have $\mathcal{A} \triangleleft f(\mathcal{B}'_1, \ldots, \mathcal{B}'_k)$ for a term operation $f$ of arity $k$ and adversaries $\mathcal{B}'_1, \ldots, \mathcal{B}'_k \in \mathcal{S}'$. And, for all $i \in [k]$ we have $\mathcal{B}'_i \triangleleft g_i(\mathcal{B}_{(i,1)}, \ldots, \mathcal{B}_{(i,m_i)})$ for a term operation $g$ of arity $m_i$ and adversaries $\mathcal{B}_{(i,1)}, \ldots, \mathcal{B}_{(i,m_i)} \in \mathcal{S}$. Define $h$ to be the term operation of arity $m = \sum_{i=1}^k m_i$ such that $h(x_1, \ldots, x_m) = f(g_1(x_1, \ldots, x_{m_1}), g_2(x_{m_1+1}, \ldots, x_{m_1+m_2}), \ldots)$. We claim that

$$\mathcal{A} \triangleleft h(\mathcal{B}_{(1,1)}, \ldots, \mathcal{B}_{(1,m_1)}, \mathcal{B}_{(2,1)}, \ldots, \mathcal{B}_{(2,m_2)}, \ldots).$$

For all $i \in [n]$, where $n$ denotes the length of the adversaries under discussion, we have

$$\begin{aligned}
\mathcal{A}_i &\subseteq f(\mathcal{B}'_{1i}, \ldots, \mathcal{B}'_{ki}) \\
&\subseteq f(g_1(\mathcal{B}_{(1,1)i}, \ldots, \mathcal{B}_{(1,m_1)i}), g_2(\mathcal{B}_{(2,1)i}, \ldots, \mathcal{B}_{(2,m_2)i}), \ldots) \\
&= h(\mathcal{B}_{(1,1)i}, \ldots, \mathcal{B}_{(1,m_1)i}, \mathcal{B}_{(2,1)i}, \ldots, \mathcal{B}_{(2,m_2)i}, \ldots).
\end{aligned}$$

The first inclusion follows from $\mathcal{A} \triangleleft f(\mathcal{B}'_1, \ldots, \mathcal{B}'_k)$, the second inclusion follows from $\mathcal{B}'_i \triangleleft g_i(\mathcal{B}_{(i,1)}, \ldots, \mathcal{B}_{(i,m_i)})$ for all $i \in [k]$, and the equality follows from the definition of $h$. $\square$

We can now show that the collapsibility results on constraint languages obtained in the previous section can in fact be interpreted as collapsibility results on algebras. For example, let us consider the proof of Theorem 4.3, which concerned constraint languages over domain $\{0,1\}$ having the boolean AND $\wedge$ as polymorphism. In that proof, we started by assuming the adversaries $\mathsf{Adv}(n, \{1\}, 1, A)$ to be $\Phi$-winnable, and repeatedly used $\wedge$-composability to derive the winnability of larger adversary sets, eventually deriving the winnability of $A^n$. By Proposition 5.3, that proof establishes that for any algebra $\mathbb{A}$ with universe $\{0,1\}$ having $\wedge$ as term operation, the adversary $A^n$ is $\mathbb{A}$-composable from the set of adversaries $\mathsf{Adv}(n, \{1\}, 1, A)$. Glancing back at Definition 5.1, we can see that the following is implied by that proof.

**Theorem 5.4** *An algebra with universe $\{0,1\}$ and having the boolean AND $\wedge$ as term operation is collapsible with source $\{1\}$ and width $1$.*

In a similar manner, the following theorems can be derived from the proofs of Theorems 4.6 and 4.9.

**Theorem 5.5** *An algebra with universe $A$ having a Mal'tsev term operation is collapsible with source $\{a\}$ and width $1$, for any $a \in A$.*

**Theorem 5.6** *An algebra with universe $A$ having the dual discriminator operation (over $A$) as term operation is collapsible with width $1$.*

The notion of collapsibility for an algebra concerns the $\mathbb{A}$-composability of the adversary $A^n$. It will be useful to consider, more generally, the $\mathbb{A}$-composability of adversaries $B^n$ for $B$ a subset of $A$.

**Definition 5.7** *A subset $B$ of $A$ is $\mathbb{A}$-collapsible with source $S \subseteq A$ and width $w \geq 0$ if for all $n \geq 1$, the adversary $B^n$ is $\mathbb{A}$-composable from the set of adversaries $\cup_{a \in S}\mathsf{Adv}(n, \{a\}, w, A)$.*

Note that, in the language of this definition, an algebra $\mathbb{A}$ is collapsible if and only if its universe $A$ is $\mathbb{A}$-collapsible.
We end this subsection with two observations.

**Observation 5.8** *Every one-element subset $B$ of $A$ is $\mathbb{A}$-collapsible with source $B$ and width $0$.*

**Observation 5.9** *If a subalgebra $\mathbb{B} = (B, F|_B)$ of $\mathbb{A} = (A, F)$ is collapsible, then $B$ is $\mathbb{A}$-collapsible.*

## 5.2 Extending Subsets

We now present a technique for establishing collapsibility results based on the notion of an operation *extending* a subset to a larger subset.

**Definition 5.10** *Let $B$ and $B'$ be subsets of $A$ with $B \subseteq B'$. We say that an operation $f : A^k \to A$ (of arity $k \geq 2$) extends $B$ to $B'$ if for every $i \in [k]$, it holds that $B' \subseteq \{f(b_1, \ldots, b_k) : b_i \in B, b_j \in B' \text{ for } j \in [k] \setminus \{i\}\}$.*

The following is the main theorem we have concerning this notion.

**Theorem 5.11** *Let $B$ and $B'$ be subsets of $A$ with $B \subseteq B'$. Suppose that $B$ is $\mathbb{A}$-collapsible with source $S$ and width $w$, and there exists an arity $k$ term operation of $\mathbb{A}$ extending $B$ to $B'$. Then, $B'$ is $\mathbb{A}$-collapsible with source $S$ and width $w + (k - 1)$.*

Before proving this theorem, we establish a lemma that will be helpful.

**Lemma 5.12** *If $B$ is $\mathbb{A}$-collapsible with source $S$ and width $w$, then for all $n \geq 1$, all adversaries in $\mathsf{Adv}(n, B, k - 1, A)$ are $\mathbb{A}$-composable from $\cup_{a \in S}\mathsf{Adv}(n, \{a\}, w + (k - 1), A)$.*

**Proof**. Let $\mathcal{A}(n, B, I, A)$ be an arbitrary adversary from $\mathsf{Adv}(n, B, k - 1, A)$. By hypothesis, there exist a term operation $g : A^m \to A$ and adversaries $\mathcal{B}_1, \ldots, \mathcal{B}_m \in \cup_{a \in S}\mathsf{Adv}(n, \{a\}, w, A)$ such that $B^n \triangleleft g(\mathcal{B}_1, \ldots, \mathcal{B}_m)$. For all $i \in [m]$, let $\mathcal{B}'_i$ be the adversary such that $\mathcal{B}'_{ij} = A$ for all $j \in I$, and $\mathcal{B}'_{ij} = \mathcal{B}_{ij}$ for all $j \in [n] \setminus I$. That is, the adversary $\mathcal{B}'_i$ is equal to the adversary $\mathcal{B}_i$, except at the coordinates in $I$ it is equal to $A$. Notice that $\mathcal{B}'_1, \ldots, \mathcal{B}'_m \in \cup_{a \in S}\mathsf{Adv}(n, \{a\}, w + (k - 1), A)$. It is straightforward to verify that $\mathcal{A}(n, B, I, A) \triangleleft g(\mathcal{B}'_1, \ldots, \mathcal{B}'_m)$. $\square$

**Proof**. (Theorem 5.11) Let $f : A^k \to A$ be an operation extending $B$ to $B'$, and fix an $n \geq 1$. We show that $B'^n$ is composable from $\cup_{a \in S}\mathsf{Adv}(n, \{a\}, w + (k - 1), A)$ in a sequence of compositions, which is justified by Proposition 5.3. By Lemma 5.12, it suffices to show that $B'^n$ is $\mathbb{A}$-composable from $\mathsf{Adv}(n, B, k - 1, A)$. Since $B' \subseteq A$, it suffices to show that $B'^n$ is $\mathbb{A}$-composable from $\mathsf{Adv}(n, B, k - 1, B')$.

We show by induction that for all $i$ such that $k-1 < i \le n$, it holds that all adversaries in $\mathsf{Adv}(n, B, i, B')$ are $\mathbb{A}$-composable from $\mathsf{Adv}(n, B, i-1, B')$. This suffices, since $B'^n \in \mathsf{Adv}(n, B, n, B')$. Let $\mathcal{A}(n, B, I, B')$ be an arbitrary adversary from $\mathsf{Adv}(n, B, i, B')$. If $|I| < i$, then $\mathcal{A}(n, B, I, B') \in \mathsf{Adv}(n, B, i-1, B')$ and the claim is obvious. So suppose that $|I| = i$. Let $d_1, \ldots, d_k$ be $k$ distinct elements from $I$, and for all $j \in [k]$, define $\mathcal{B}_j$ to be the adversary $\mathcal{A}(n, B, I \setminus \{d_j\}, A)$. We claim that $\mathcal{A}(n, B, I, B') \triangleleft f(\mathcal{B}_1, \ldots, \mathcal{B}_k)$. For all $j \notin I$, we have $\mathcal{A}(n, B, I, B')_j \subseteq f(\mathcal{B}_{1j}, \ldots, \mathcal{B}_{kj})$ because $B \subseteq f(B, \ldots, B)$, and for $j \in I$, we have $\mathcal{A}(n, B, I, B')_j \subseteq f(\mathcal{B}_{1j}, \ldots, \mathcal{B}_{kj})$ because $f$ extends $B$ to $B'$. □

A tool that will help us apply Theorem 5.11 is the following lemma.

**Lemma 5.13** *Let $f : A^k \to A$ be an operation and $a \in A$ an element. Suppose that the operation $f$ extends $\{a\}$ to $A$. Then, an algebra with universe $A$ having $f$ as term operation is collapsible with source $\{a\}$ and width $k - 1$.*

**Proof.** By Observation 5.8, the subset $\{a\}$ is $\mathbb{A}$-collapsible with source $\{a\}$ and width $0$; by Theorem 5.11, the subset $A$ is $\mathbb{A}$-collapsible with source $\{a\}$ and width $k - 1$. □

We now derive two collapsibility results which illustrate the use of Lemma 5.13.

When $f : A \times A \to A$ is a binary operation, let us say that $u \in A$ is a *unit element* of $f$ if for all $a \in A$, it holds that $f(u, a) = f(a, u) = a$.

**Theorem 5.14** *An algebra with universe $A$ having an idempotent binary term operation $f$ with unit element $u$ is collapsible with source $\{u\}$ and width $1$.*

**Proof.** It is straightforward to verify that $f$ extends $\{u\}$ to $A$; the theorem follows from Lemma 5.13. □

**Definition 5.15** *A near-unanimity operation is an operation $f : A^k \to A$ of arity $k \ge 3$ satisfying the identities $x = f(y, x, x, \ldots, x) = f(x, y, x, \ldots, x) = \cdots = f(x, \ldots, x, x, y)$. In words, if all but at most one of the arguments of $f$ are equal to $x$, then $x$ is the result of the operation.*

**Theorem 5.16** *An algebra with universe $A$ having a near-unanimity term operation is collapsible with source $\{a\}$ and width $k - 1$, for any $a \in A$.*

**Proof.** It is straightforward to verify that $f$ extends $\{a\}$ to $A$ for any $a \in A$; the theorem follows from Lemma 5.13. □

There is overlap between the results that we just gave and the collapsibility results of the previous section:

- Theorem 5.14 implies Theorem 4.3, since the boolean AND $\wedge$ has $1$ as unit element.

- Theorems 5.16 and 4.9 overlap: a dual discriminator operation is an example of a near-unanimity operation, so Theorem 5.16 shows the collapsibility of a larger class of algebras. On the other hand, while applying Theorem 5.16 to a dual discriminator operation would give collapsibility with width $2$, Theorem 4.9 shows collapsibility with width $1$.

- Lemma 5.13 can also be used to derive the collapsibility of an algebra having a Mal'tsev term operation, but with a width of $2$–in contrast to the width $1$ obtained by Theorem 4.6.

We believe that it is worth highlighting that Lemma 5.13 permits us to derive *in a uniform manner* all of the collapsibility results discussed thus far, although specialized arguments such as those given in the previous section may give tighter width bounds.

We now collect together the collapsibility results that can be stated for the two-element case and present them as a theorem; this will be useful in our study of the three-element case. Define a binary operation $f : A \times A \to A$ to be a *semilattice operation* if it is associative, commutative, and idempotent. Note that every semilattice operation over a two-element domain $A = \{a, b\}$ has a unit element, namely, the single element contained in $A \setminus \{f(a, b), f(b, a)\}$. As examples, the boolean AND operation $\wedge$ and boolean OR operation $\vee$ are the two semilattice operations over the two-element domain $\{0, 1\}$, having unit elements $1$ and $0$, respectively.

**Theorem 5.17** *Suppose that $\mathbb{A}$ is a two-element idempotent algebra that is not a G-set. Then $\mathbb{A}$ is collapsible with a one-element source. In particular, at least one of the following holds.*

- *The algebra $\mathbb{A}$ contains a semilattice term operation $f$, and the unit element of $f$ serves as a source.*

- *The algebra $\mathbb{A}$ contains a near-unanimity term operation, and either element of $A$ serves as a source.*

- *The algebra $\mathbb{A}$ contains the operation $\text{minority}_A$ as a term operation, and either element of $A$ serves as a source.*

**Proof**. That the algebra $\mathbb{A}$ contains one of the named operations follows from Theorem 2.25. The collapsibility results with the claimed sources follow from Theorems 5.14, 5.16, and 4.6, respectively. □

## 5.3 From Subsets to Subalgebras

Here, we demonstrate that subsets appearing in an adversary can be enlarged to subalgebras. This is made precise in the following proposition.

**Proposition 5.18** *Suppose that an adversary $(A_1, \ldots, A_n)$ is $\mathbb{A}$-composable from a set of adversaries $\mathcal{S}$, and let $B_i$ denote the subalgebra of $\mathbb{A}$ generated by $A_i$ for all $i \in [n]$. The adversary $(B_1, \ldots, B_n)$ is $\mathbb{A}$-composable from $\mathcal{S}$.*

**Proof**. Assume that the adversary $(A_1, \ldots, A_n)$ is $\mathbb{A}$-composable from $\mathcal{S}$, and that $A_k$ is not a subalgebra of $\mathbb{A}$, where $k \in [n]$. It suffices to show that an adversary $(A'_1, \ldots, A'_n)$ with $A'_i \supseteq A_i$ for all $i \in [n]$ and $A'_k \supsetneq A_k$, is $\mathbb{A}$-composable from $\mathcal{S}$; iteratively applying this result gives the proposition.

Since $A_k$ is not a subalgebra of $\mathbb{A}$, there exists an operation $f$ of $\mathbb{A}$ such that $f(A_k, \ldots, A_k)$ contains an element outside of $A_k$. Define $A'_i$ by $f(A_i, \ldots, A_i)$ for all $i \in [n]$. We have

$$(A'_1, \ldots, A'_n) \triangleleft f((A_1, \ldots, A_n), \ldots, (A_1, \ldots, A_n))$$

and so $(A'_1, \ldots, A'_n)$ is $\mathbb{A}$-composable from $\{(A_1, \ldots, A_n)\}$. By Proposition 5.3, $(A'_1, \ldots, A'_n)$ is $\mathbb{A}$-composable from $\mathcal{S}$. The adversary $(A'_1, \ldots, A'_n)$ has the desired properties: we have $A'_i \supseteq A_i$ for all $i \in [n]$ by the idempotence of $f$, and $A'_k = f(A_k, \ldots, A_k)$ contains an element outside of $A_k$. □

Proposition 5.18 has the following consequence.

**Proposition 5.19** *Let $B$ be a subset of $A$ that is $\mathbb{A}$-collapsible with source $S$. The subalgebra generated by $B$ is also $\mathbb{A}$-collapsible with source $S$.*

**Proof**. Suppose that $B$ is $\mathbb{A}$-collapsible with source $S$ and width $w \geq 0$. For each $n \geq 1$, we have that $B^n$ is $\mathbb{A}$-composable from $\cup_{a \in S} \text{Adv}(n, \{a\}, w, A)$. Let $B'$ denote the subalgebra of $\mathbb{A}$ generated by $B$. By Proposition 5.18, the adversary $B'^n$ is $\mathbb{A}$-composable from $\cup_{a \in S} \text{Adv}(n, \{a\}, w, A)$, and so $B'$ is $\mathbb{A}$-collapsible with source $S$. □

As an application of Proposition 5.19, we can observe the collapsibility of a certain class of "basic" algebras. A finite algebra $\mathbb{A}$ is *simple* if any homomorphic image of $\mathbb{A}$ smaller than $\mathbb{A}$ is one-element; and, is *strictly simple* if it is simple and all of of its proper subalgebras are one-element.

**Theorem 5.20** *Let $\mathbb{A}$ be a strictly simple finite idempotent algebra. If $\mathbb{A}$ does not contain a G-set as factor, then it is collapsible with a one-element source.*

Strictly simple algebras were studied in the context of CSP complexity by Bulatov, Jeavons, and Krokhin [7]. To establish Theorem 5.20, we make use of a result from that work.

**Theorem 5.21** *(follows from [7, proof of Theorem 6.2]) Let $\mathbb{A}$ be a strictly simple finite idempotent algebra. Either $\mathbb{A}$ is a G-set, or contains a term operation of one of the following types:*

- *a dual discriminator,*
- *a Mal'tsev operation,*
- *a semilattice operation $f : A \times A \to A$ of the form*

$$f(x,y) = \begin{cases} x & \text{if } x = y \\ m & \text{otherwise} \end{cases}$$

*for an element $m \in A$.*

**Proof**. (Theorem 5.20) Suppose that $\mathbb{A}$ does not have a G-set as factor. Then it must contain a term operation of one of the three types given in the statement of Theorem 5.21. If $\mathbb{A}$ contains a dual discriminator operation or a Mal'tsev operation, it is collapsible with a one-element source by Theorem 4.9 or 4.6, respectively. So suppose that $\mathbb{A}$ contains a semilattice operation $f$ of the described form. Fix $a \in A$ to be an element distinct from $m$. By Observation 5.8, the set $\{a\}$ is $\mathbb{A}$-collapsible with source $\{a\}$. Now observe that $f$ extends $\{a\}$ to $\{a, m\}$; hence, by Theorem 5.11, the set $\{a, m\}$ is $\mathbb{A}$-collapsible with source $\{a\}$. Since $\mathbb{A}$ is strictly simple, the smallest subalgebra of $\mathbb{A}$ containing $\{a, m\}$ is $\mathbb{A}$ itself, and so $\mathbb{A}$ is collapsible with source $\{a\}$ by Proposition 5.19. □

## 5.4 Combining Subsets

We now demonstrate that the collapsibility of the union $B_1 \cup B_2$ of two subsets can be inferred from the collapsibility of the two subsets individually, along with an assumption stating that the source of one subset must fall into the other.

**Theorem 5.22** *Let $B_1, B_2$ be subsets of $A$ such that*

- *$B_1$ is $\mathbb{A}$-collapsible with source $S_1$ and width $w_1$, and*
- *$B_2$ is $\mathbb{A}$-collapsible with source $S_2$ where $S_2 \subseteq B_1$, and width $w_2$.*

*Then, $B_1 \cup B_2$ is $\mathbb{A}$-collapsible with source $S_1$ and width $w_1 + w_2$.*

**Proof**. Let $n \geq 1$. By hypothesis, we have $B_1^n \triangleleft f(\mathcal{A}(n, \{s_1\}, I_1, A), \ldots, \mathcal{A}(n, \{s_k\}, I_k, A))$ for $f$ a term operation of $\mathbb{A}$, elements $s_1, \ldots, s_k \in S_1$, and subsets $I_1, \ldots, I_k \subseteq [n]$ with $|I_1|, \ldots, |I_k| \leq w_1$.

We claim that every adversary in $\mathsf{Adv}(n, B_1, w_2, A)$ is $\mathbb{A}$-composable from $\cup_{s \in S_1} \mathsf{Adv}(n, \{s\}, w_1 + w_2, A)$. Consider an adversary $\mathcal{A}(n, B_1, I, A)$ with $I \subseteq [n]$, $|I| \leq w_2$, that is, an arbitrary adversary from $\mathsf{Adv}(n, B_1, w_2, A)$. Observe that each of the adversaries $\mathcal{A}(n, \{s_1\}, I_1 \cup I, A), \ldots, \mathcal{A}(n, \{s_k\}, I_k \cup I, A)$ is in the set $\cup_{s \in S_1} \mathsf{Adv}(n, \{s\}, w_1 + w_2, A)$. We show that

$$\mathcal{A}(n, B_1, I, A) \triangleleft f(\mathcal{A}(n, \{s_1\}, I_1 \cup I, A), \ldots, \mathcal{A}(n, \{s_k\}, I_k \cup I, A)).$$

We verify
$$\mathcal{A}(n, B_1, I, A)_j \subseteq f(\mathcal{A}(n, \{s_1\}, I_1 \cup I, A)_j, \ldots, \mathcal{A}(n, \{s_k\}, I_k \cup I, A)_j)$$
for all $j \in [n]$ as follows:

- For $j \in I$, we have $A \subseteq f(A, \ldots, A)$ by the idempotence of $f$.

- For $j \in [n] \setminus I$, it holds that $\mathcal{A}(n, \{s_i\}, I_i, A)_j = \mathcal{A}(n, \{s_i\}, I_i \cup I, A)$, and the containment follows from $B_1^n \triangleleft f(\mathcal{A}(n, \{s_1\}, I_1, A), \ldots, \mathcal{A}(n, \{s_k\}, I_k, A))$.

We have shown that every adversary in $\mathsf{Adv}(n, B_1, w_2, A)$ is $\mathbb{A}$-composable from $\cup_{s \in S_1} \mathsf{Adv}(n, \{s\}, w_1 + w_2, A)$. By appeal to Proposition 5.3, it suffices to show that $(B_1 \cup B_2)^n$ is $\mathbb{A}$-composable from $\mathsf{Adv}(n, B_1, w_2, A)$. By hypothesis, we have $B_2^n \triangleleft g(\mathcal{A}(n, \{t_1\}, J_1, A), \ldots, \mathcal{A}(n, \{t_m\}, J_m, A))$ for $g$ a term operation of $\mathbb{A}$, elements $t_1, \ldots, t_m \in S_2$, and subsets $J_1, \ldots, J_m \subseteq [n]$ with $|J_1|, \ldots, |J_m| \leq w_2$. We claim that $(B_1 \cup B_2)^n \triangleleft g(\mathcal{A}(n, B_1, J_1, A), \ldots, \mathcal{A}(n, B_1, J_m, A))$. We verify this as follows. For each $j \in [n]$, we have $B_1 \subseteq g(B_1, \ldots, B_1) \subseteq g(\mathcal{A}(n, B_1, J_1, A)_j, \ldots, \mathcal{A}(n, B_1, J_m, A)_j)$; the first containment follows from the idempotence of $g$, and the second containment follows from the fact that $B_1 \subseteq \mathcal{A}(n, B_1, J_i, A)_j$ for all $i \in [m]$. Also, for each $j \in [n]$, we have $B_2 \subseteq g(\mathcal{A}(n, \{t_1\}, J_1, A)_j, \ldots, \mathcal{A}(n, \{t_m\}, J_m, A)_j) \subseteq g(\mathcal{A}(n, B_1, J_1, A)_j, \ldots, \mathcal{A}(n, B_1, J_m, A)_j)$; the first containment was given above, and the second containment follows from the fact that $\mathcal{A}(n, \{t_i\}, J_i, A)_j \subseteq \mathcal{A}(n, B_1, J_i, A)_j$ for all $i \in [m]$, as $\{t_i\} \subseteq B_1$ for all $i \in [m]$. $\square$

Now we give an application of Theorem 5.22. Let us say that an algebra $\mathbb{A}$ having two or more elements is *pair minimal* if for every two-element subset $B$ of $A$, the smallest subalgebra of $\mathbb{A}$ containing $B$ is minimal in that it does not properly contain any non-trivial subalgebras.

**Theorem 5.23** *If $\mathbb{A}$ is a pair minimal algebra that does not have a G-set as factor, then $\mathbb{A}$ is collapsible.*

**Proof**. We show that every subset $B \subseteq A$ is $\mathbb{A}$-collapsible with a one-element source, by induction on $|B|$. For subsets $B$ with $|B| = 1$, this is immediate from Observation 5.8.

Suppose that $B \subseteq A$ is $\mathbb{A}$-collapsible with source $\{s\}$ and suppose $a \in A$. We show that there is a subset of $A$ containing $B \cup \{a\}$ that is $\mathbb{A}$-collapsible with a one-element source. Let $B'$ be the smallest subalgebra of $\mathbb{A}$ containing $\{s, a\}$. Because $\mathbb{A}$ is pair minimal, $B'$ does not properly contain any non-trivial subalgebras. By Fact 2.16, it follows that any homomorphic image of $\mathbb{A}$ smaller than $\mathbb{A}$ is one-element, and thus $B'$ is a strictly simple algebra. By Theorem 5.20, and Observation 5.9, the set $B'$ is $\mathbb{A}$-collapsible with a one-element source $\{s'\}$. By Theorem 5.22 with $B_1 = B'$, $S_1 = \{s'\}$, $B_2 = B$, and $S_2 = \{s\}$, we have that $B \cup B'$ is $\mathbb{A}$-collapsible with source $\{s'\}$. $\square$

Any algebra $\mathbb{A} = (A, F)$ where every two-element subset $B$ of $A$ is a subalgebra can immediately be seen to be pair minimal. We can therefore derive the following corollary from Theorem 5.23.

**Corollary 5.24** *If $\mathbb{A}$ is an algebra that does not have a G-set as factor where every two-element subset $B$ of $A$ is a subalgebra of $\mathbb{A}$, then $\mathbb{A}$ is collapsible.*

From this corollary and Bulatov's Theorem 2.24, we can derive the following result.

**Corollary 5.25** *Let $\Gamma$ be a constraint language over $A$ containing all subsets of $A$. Then $\mathsf{QCSP}_\mathsf{c}(\Gamma)$ is in P if the algebra $\mathbb{A}_\Gamma$ does not have a G-set as factor, and NP-hard otherwise.*

**Proof**. If the algebra $\mathbb{A}_\Gamma$ has a G-set as factor, then $\mathsf{QCSP}_\mathsf{c}(\Gamma)$ is NP-hard by Theorem 2.22. If the algebra $\mathbb{A}_\Gamma$ does not have a G-set as factor, then $\mathbb{A}_\Gamma$ is collapsible by Corollary 5.24; note that, since all subsets of $A$ are in $\Gamma$, all subsets of $A$ are subalgebras of $\mathbb{A}_\Gamma$. By Proposition 5.2, $\mathsf{QCSP}_\mathsf{c}(\Gamma)$ reduces to $\mathsf{CSP}_\mathsf{c}(\Gamma)$, and by Theorem 2.24, the problem $\mathsf{CSP}_\mathsf{c}(\Gamma)$ is in P. $\square$

# 6 Sink Algebras

In this section, we identify a class of algebras that we call *sinks*. We prove a theorem which shows that any algebra must have as factor a sink or a G-set, or has the desirable property of being collapsible. Now, any constraint language whose algebra has a G-set as factor is known to be hard (Theorem 2.22); and, for any constraint language $\Gamma$ whose algebra $\mathbb{A}_\Gamma$ is collapsible the problem $\mathsf{QCSP_c}(\Gamma)$ has the same complexity as $\mathsf{CSP_c}(\Gamma)$ (Proposition 5.2). Therefore, this theorem effectively reduces the classification of tractable problems $\mathsf{QCSP_c}(\Gamma)$ to 1) the study of sinks, and 2) the classification of the problems $\mathsf{CSP_c}(\Gamma)$. In the next section, we will demonstrate the utility of both the definition of *sink* and the accompanying theorem, where they will be used to study the three-element case.

We need to introduce two properties of algebras before defining the notion of a *sink*.

**Definition 6.1** *A finite idempotent algebra $\mathbb{A}$ is* enclosed *if for every term operation $f : A^k \to A$ of $\mathbb{A}$ and maximal proper subalgebras $B_1, \ldots, B_k$ of $\mathbb{A}$, there exists a maximal proper subalgebra $B$ such that $f(B_1, \ldots, B_k) \subseteq B$.*

Let $\mathbb{A}$ be a finite idempotent algebra. When $B, B'$ are two maximal proper subalgebras of $\mathbb{A}$, we say that $B$ overlaps $B'$ (or, that $B$ and $B'$ overlap) if $B \cap B' \neq \emptyset$; and, we say that $B$ and $B'$ are *connected* if there exists a sequence of maximal proper subalgebras $B_1, \ldots, B_k$ such that $B = B_1$, $B' = B_k$, and $B_i$ overlaps $B_{i+1}$ for all $i \in [k-1]$.

**Definition 6.2** *A finite idempotent algebra $\mathbb{A}$ is* fully connected *if all pairs of maximal proper subalgebras of $\mathbb{A}$ are connected.*

We can now give the definition of a *sink algebra*.

**Definition 6.3** *A finite idempotent algebra $\mathbb{A} = (A, F)$ is a* sink *if it is enclosed, fully connected, does not contain a G-set as factor, and is not collapsible.*

We now prove the promised theorem concerning sinks. This theorem shows that an algebra is collapsible so long as it excludes two types of "forbidden" algebras: sinks and G-sets.

**Theorem 6.4** *If $\mathbb{A} = (A, F)$ is a finite idempotent algebra that does not contain a sink nor a G-set as factor, then $\mathbb{A}$ is collapsible.*

**Proof**. We prove this theorem by induction on $|A|$, the size of the universe of the algebra $\mathbb{A}$. It is trivial for $|A| = 1$, by Observation 5.8. If $\mathbb{A}$ has no non-trivial proper subalgebra, then any homomorphic image of $\mathbb{A}$ smaller than $\mathbb{A}$ must be one-element by Fact 2.16, and hence $\mathbb{A}$ is strictly simple; in this case, the theorem follows from Theorem 5.20. We therefore assume that $\mathbb{A}$ has a non-trivial proper subalgebra.

If the algebra $\mathbb{A}$ is not enclosed, then, by definition, there exists a term operation $f : A^k \to A$ of $\mathbb{A}$ and maximal proper subalgebras $B_1, \ldots, B_k$ of $\mathbb{A}$ such that the smallest subalgebra of $\mathbb{A}$ containing $f(B_1, \ldots, B_k)$ is $\mathbb{A}$ itself. By induction and Observation 5.9, we may assume that each of the sets $B_1, \ldots, B_k$ are $\mathbb{A}$-collapsible. Because $f(B_1, \ldots, B_k)^n \triangleleft f(B_1^n, \ldots, B_k^n)$ for all $n \geq 1$, we have that $f(B_1, \ldots, B_k)$ is $\mathbb{A}$-collapsible; by Proposition 5.19 with $B = f(B_1, \ldots, B_k)$, we have that $A$ is $\mathbb{A}$-collapsible. Let us therefore assume for the rest of the proof that the algebra $\mathbb{A}$ is enclosed.

The following lemma provides some structural information concerning $\mathbb{A}$.

**Lemma 6.5** *Suppose that $\mathbb{A}$ is a finite idempotent algebra that is enclosed. Then, $\mathbb{A}$ is either fully connected, or its maximal proper subalgebras are disjoint.*

**Proof**. Let $\mathbf{B} = \{B_i\}_{i \in I}$ be a collection of maximal proper subalgebras that is closed in the sense that if a maximal proper subalgebra $B$ is connected to a $B_i$ in the collection, then $B$ is in the collection. It suffices to show that $\cup_{i \in I} B_i$ is in fact a subalgebra of $\mathbb{A}$. We will establish the following claim.

**Claim:** Let $f : A^k \to A$ be a term operation of $\mathbb{A}$, let $B_1, \ldots, B_k$ be subalgebras from $\mathbf{B}$, and let $B'_1, \ldots, B'_k$ be subalgebras from $\mathbf{B}$ such that $B_i$ overlaps $B'_i$ for all $i \in [k]$. If $B$ is a maximal proper subalgebra such that $f(B_1, \ldots, B_k) \subseteq B$, then there exists a maximal proper subalgebra $B'$ overlapping $B$ such that $f(B'_1, \ldots, B'_k) \subseteq B'$.

We can use the claim to show that $\cup_{i \in I} B_i$ is a subalgebra, as follows. Let $f : A^k \to A$ be an arbitrary term operation of $\mathbb{A}$. Let $i_1, \ldots, i_k \in I$ be arbitrary. Fix $B_0$ to be any member of $\mathbf{B}$. We have $f(B_0, \ldots, B_0) \subseteq B_0$. Since $B_0$ is connected to each of $B_{i_1}, \ldots, B_{i_k}$, by repeated application of the claim, we can establish that $f(B_{i_1}, \ldots, B_{i_k})$ is contained in a maximal proper subalgebra connected to $B_0$, implying that $f(B_{i_1}, \ldots, B_{i_k})$ is contained in a member of $\mathbf{B}$. Since the indices $i_1, \ldots, i_k$ were arbitrary, this implies that $\cup_{i \in I} B_i$ is a subalgebra of $\mathbb{A}$.

We now prove the claim. We have $f(B_1, \ldots, B_k) \subseteq B$. It follows that $f(B_1 \cap B'_1, \ldots, B_k \cap B'_k) \subseteq B$. Since $f(B_1 \cap B'_1, \ldots, B_k \cap B'_k) \subseteq f(B'_1, \ldots, B'_k)$, we have that $B$ overlaps $f(B'_1, \ldots, B'_k)$. Since $\mathbb{A}$ is enclosed, $f(B'_1, \ldots, B'_k)$ is contained in a maximal proper subalgebra $B'$. From the facts that $B$ overlaps $f(B'_1, \ldots, B'_k)$ and that $f(B'_1, \ldots, B'_k) \subseteq B'$, we have that $B$ overlaps $B'$, giving the claim. $\square$

By Lemma 6.5, we have that $\mathbb{A}$ is either fully connected, or that its maximal proper subalgebras are disjoint.

Suppose that $\mathbb{A}$ is fully connected. Then, we have that $\mathbb{A}$ is fully connected, is enclosed, and (by hypothesis) does not contain a G-set as factor; also by hypothesis, $\mathbb{A}$ is not a sink, so by definition of a sink, we have that $\mathbb{A}$ is collapsible.

Suppose that the maximal proper subalgebras of $\mathbb{A}$ are disjoint. In this case, the following lemma shows the collapsibility of $\mathbb{A}$.

**Lemma 6.6** *Suppose that $\mathbb{A}$ is a finite idempotent algebra that is enclosed and such that the maximal proper subalgebras of $\mathbb{A}$ are disjoint. Then the equivalence relation $\theta \subseteq A \times A$ having the maximal proper subalgebras of $\mathbb{A}$ as its equivalence classes is a congruence. And, if the homomorphic image of $\mathbb{A}$ given by $\theta$ is collapsible with source $S$, then $\mathbb{A}$ is collapsible with any source $T$ such that $S \subseteq \{a^\theta : a \in T\}$.*

**Proof.** Let $\theta$ be the equivalence relation having the maximal proper subalgebras of $\mathbb{A}$ as its equivalence classes. We first verify that $\theta$ is a congruence. We need to show that each operation $f \in F$ is a polymorphism of $\theta$. Let $f$ be an arity $k$ operation from $F$, and suppose $(b_1, b'_1), \ldots, (b_k, b'_k) \in \theta$. We have to demonstrate that $(f(b_1, \ldots, b_k), f(b'_1, \ldots, b'_k)) \in \theta$. By definition of $\theta$, for all $i \in [k]$, there exists a maximal proper subalgebra $B_i$ such that $b_i, b'_i \in B_i$. Since $\mathbb{A}$ is enclosed, there exists a maximal proper subalgebra $B$ such that $f(B_1, \ldots, B_k) \subseteq B$. Thus $f(b_1, \ldots, b_k), f(b'_1, \ldots, b'_k) \in B$, and $(f(b_1, \ldots, b_k), f(b'_1, \ldots, b'_k)) \in \theta$.

Now assume that the homomorphic image $(A^\theta, F^\theta)$ of $\mathbb{A}$ is collapsible with source $S$ and width $w \geq 0$. Assume also that $T$ is a subset of $A$ such that $S \subseteq \{a^\theta : a \in T\}$. We claim that $\mathbb{A}$ is collapsible with source $T$ and width $w$.

Let $n \geq 1$. There exists an operation $f \in F$ such that $(A^\theta)^n \triangleleft f^\theta(\mathcal{B}_1, \ldots, \mathcal{B}_k)$ where $\mathcal{B}_1, \ldots, \mathcal{B}_k \in \cup_{s \in S} \mathsf{Adv}(n, \{s\}, w, A^\theta)$. For all $i \in [k]$, define the element $s_i$ to be an element of $S$ such that $\mathcal{B}_i \in \mathsf{Adv}(n, \{s_i\}, w, A^\theta)$. Let $t_i$ be an element in $T$ such that $s_i = t_i^\theta$. For all $i \in [k]$, define $\mathcal{B}'_i \in \mathsf{Adv}(n, \{t_i\}, w, A)$. to be the adversary such that

$$\mathcal{B}'_{ij} = \begin{cases} A & \text{if } \mathcal{B}_{ij} = A^\theta \\ \{t_i\} & \text{if } \mathcal{B}_{ij} = \{s_i\}. \end{cases}$$

We have that $A^\theta \subseteq f^\theta(\mathcal{B}_{1j}, \ldots, \mathcal{B}_{kj})$ for all $j \in [n]$. Observe that $\mathcal{B}_{ij} = (\mathcal{B}'_{ij})^\theta$ for all $i \in [k]$ and $j \in [n]$. Thus $A^\theta \subseteq f^\theta(\mathcal{B}_{1j}, \ldots, \mathcal{B}_{kj}) \subseteq f^\theta((\mathcal{B}'_{1j})^\theta, \ldots, (\mathcal{B}'_{kj})^\theta) \subseteq (f(\mathcal{B}'_{1j}, \ldots, \mathcal{B}'_{kj}))^\theta$. Letting $A_j$ denote $f(\mathcal{B}'_{1j}, \ldots, \mathcal{B}'_{kj})$, we have $A^\theta \subseteq (A_j)^\theta$ and $(A_1, \ldots, A_n) \triangleleft f(\mathcal{B}'_1, \ldots, \mathcal{B}'_k)$. Since for each $j \in [n]$ we have the containment $A^\theta \subseteq (A_j)^\theta$, the set $A_j$ contains one element from each equivalence class of $\theta$, and hence the subalgebra generated by $A_j$ is $A$ itself. By Proposition 5.18, we conclude that $A^n$ is $\mathbb{A}$-composable from $\cup_{t \in T} \mathsf{Adv}(n, \{t\}, w, A)$. □

Note that the homomorphic image of $\mathbb{A}$ given by $\theta$ is smaller than $\mathbb{A}$, as $\mathbb{A}$ has a non-trivial proper subalgebra. This homomorphic image is thus collapsible by induction, and we can apply Lemma 6.6 to derive the collapsibility of $\mathbb{A}$. □

# 7 The Three-Element Case

This section uses the ideas developed throughout this paper to investigate the complexity of $\mathsf{QCSP}_\mathsf{c}(\Gamma)$ for constraint languages over a three-element domain. In particular, we analyze three-element sink algebras, showing that any such algebra must have a particular semilattice operation as term operation. This result will allow us to establish a classification of $\mathsf{QCSP}_\mathsf{c}(\Gamma)$ for all constraint languages $\Gamma$ that do not have the identified semilattice operation as polymorphism.

We begin by observing that there are no one- nor two-element sinks.

**Observation 7.1** *There are no one- nor two-element sinks.*

**Proof**. By definition, a sink does not contain a G-set as factor, and is not collapsible. Observation 5.8 implies that any one-element algebra is collapsible, and Theorem 5.17 shows that any two-element algebra not containing a G-set as factor is collapsible. □

We will show, however, that there are three-element sinks. We begin our investigation of three-element sinks by showing that any such sink must contain a particular semilattice operation.

**Theorem 7.2** *Let $\mathbb{A} = (A, F)$ be an three-element idempotent algebra that is a sink. Then, $\mathbb{A}$ has exactly two subalgebras of size two. Let $c$ denote the common element of these two subalgebras, let $a, b$ denote the other two elements, and let $s_{abc} : A \times A \to A$ denote the semilattice operation defined by $s_{abc}(x, y) = c$ if $x \neq y$, and $s_{abc}(x, y) = x$ if $x = y$. The algebra $\mathbb{A}$ has $s_{abc}$ as a term operation.*

**Proof**. Suppose that $\mathbb{A}$ is a sink having three elements. Since a sink is fully connected by definition, each element of $A$ must be contained in a proper subalgebra of size strictly greater than one. Hence, each element of $A$ is contained in a subalgebra of size two, from which it follows that there are either two or three subalgebras of size two. If there are three subalgebras of size two, then by Corollary 5.24, the algebra $\mathbb{A}$ is collapsible, contradicting that $\mathbb{A}$ is a sink. We now have that $\mathbb{A}$ contains exactly two subalgebras. Let us denote these two subalgebras by $\alpha = \{a, c\}$ and $\beta = \{b, c\}$.

By Theorem 5.17, each of the subalgebras $\alpha$ and $\beta$ are collapsible with a one-element source. If either one of them is collapsible with source $\{c\}$, then by Theorem 5.22, we have that $\alpha \cup \beta = A$ is $\mathbb{A}$-collapsible (that is, $\mathbb{A}$ is collapsible), contradicting that $\mathbb{A}$ is a sink. Hence, neither of $\alpha, \beta$ is collapsible with source $\{c\}$; by Theorem 5.17, we have that

- the subalgebra $\alpha$ contains a semilattice term operation $s_\alpha$ with $a$ as unit element,

- the subalgebra $\beta$ contains a semilattice term operation $s_\beta$ with $b$ as unit element, and

- neither of the subalgebras $\alpha, \beta$ contains a semilattice term operation with $c$ as the unit element.

It follows that $\mathbb{A}$ contains a term operation $s_a$ whose restriction to $\{a, c\}$ is $s_\alpha$, and a term operation $s_b$ whose restriction to $\{b, c\}$ is $s_\beta$.

Observe that, for every binary term operation $f$ of $\mathbb{A}$, we cannot have $f(a, c) = f(c, a) = a$, otherwise the subalgebra $\alpha$ would contain a semilattice term operation with $c$ as the unit element. Therefore, the restriction of $f$ to $\{a, c\}$ is either a projection (when $f(a, c) \neq f(c, a)$), or the semilattice operation $s_a$ (when $f(a, c) = f(c, a) = c$). Likewise, the restriction of $f$ to $\{b, c\}$ is either a projection, or the semilattice operation $s_b$. From these observations, it is straightforward to verify that the binary term operation $s' : A \times A \to A$ defined by $s'(x, y) = s_b(s_a(x, y), s_a(y, x))$ is equal to $s_{abc}$, except possibly at the points $(a, b), (b, a)$. Because $\{a, b\}$ is not a subalgebra of $\mathbb{A}$, there exists a binary term operation $r : A \times A \to A$ of $\mathbb{A}$ with $r(a, b) = c$. Using the observations again for $r$, it is straightforward to verify that the binary term operation $s'' : A \times A \to A$ defined by $s''(x, y) = s'(r(x, y), r(y, x))$ is equal to $s_{abc}$. □

By combining this result with the previous section's theorem on sinks and Bulatov's Theorem 2.23, we can state a $\mathsf{QCSP_c}(\Gamma)$ tractability classification of all constraint languages $\Gamma$, over a three-element domain, that do not have the identified semilattice polymorphism.

**Theorem 7.3** *Let $\Gamma$ be a constraint language over a three-element domain $D$. Suppose that there is no way to label the elements of $D$ as $a, b, c$ such that $s_{abc}$ (the operation defined in the statement of Theorem 7.2) is a polymorphism of $\Gamma$. Then, the problem $\mathsf{QCSP_c}(\Gamma)$ is in P if the algebra $\mathbb{A}_\Gamma$ does not have a G-set as factor, and NP-hard otherwise.*

**Proof**. If the algebra $\mathbb{A}_\Gamma$ has a G-set as factor, then it is NP-hard, by Theorem 2.22. So suppose that the algebra $\mathbb{A}_\Gamma$ does not have a G-set as factor. Because $s$ is not a polymorphism of $\Gamma$, by Theorem 7.2, $\mathbb{A}_\Gamma$ is not a sink. Moreover, it does not contain a sink as factor, by Observation 7.1. By Theorem 6.4, the algebra $\mathbb{A}_\Gamma$ is collapsible. By Proposition 5.2 $\mathsf{QCSP_c}(\Gamma)$ reduces to $\mathsf{CSP_c}(\Gamma)$; and, by Bulatov's Theorem 2.23, $\mathsf{CSP_c}(\Gamma)$ is in P. □

We have shown that any three-element sink must have $s_{abc}$ as a polymorphism, but we have not yet demonstrated that any three-element sinks exist! We now give a family of examples of three-element sinks, which includes the algebra $(\{a, b, c\}, \{s_{abc}\})$ as a member.

Let us introduce some concepts and terminology. Suppose that $\mathbb{A}$ is a three-element sink. We have seen (Theorem 7.2) that $\mathbb{A}$ must have two subalgebras of size two. Let us denote these two subalgebras by $\alpha = \{a, c\}$ and $\beta = \{b, c\}$. Let us say that a term operation $f : A^k \to A$ of $\mathbb{A}$ can be *realized* as an operation $g : \{\alpha, \beta\}^k \to \{\alpha, \beta\}$ if for all $S_1, \ldots, S_k \in \{\alpha, \beta\}$, it holds that $f(S_1, \ldots, S_k) \subseteq g(S_1, \ldots, S_k)$. Notice that because $\mathbb{A}$ is fully enclosed, every term operation $f : A^k \to A$ of $\mathbb{A}$ can be realized as some operation $g : \{\alpha, \beta\}^k \to \{\alpha, \beta\}$. However, it is certainly possible that a term operation of $\mathbb{A}$ can be realized as more than one operation. For example, let us consider the operation $s_{abc}$. The following containments hold:

$$s_{abc}(\alpha, \alpha) \subseteq \alpha$$

$$s_{abc}(\alpha, \beta) \subseteq \{c\}$$

$$s_{abc}(\beta, \alpha) \subseteq \{c\}$$

$$s_{abc}(\beta, \beta) \subseteq \beta$$

Since $\{c\}$ is contained in both $\alpha$ and $\beta$, the operation $s_{abc}$ can be realized as any operation $g : \{\alpha, \beta\} \times \{\alpha, \beta\} \to \{\alpha, \beta\}$ such that $g(\alpha, \alpha) = \alpha$ and $g(\beta, \beta) = \beta$, that is, any idempotent binary operation on $\{\alpha, \beta\}$. Let us say that a term operation $f$ of $\mathbb{A}$ is $\alpha\beta$-*projective* if it can be realized by an operation $g$ on $\{\alpha, \beta\}$ that is a projection. As an example, $s_{abc}$ can be realized as either of the two binary projections over $\{\alpha, \beta\}$, and is hence $\alpha\beta$-projective.

The following theorem gives a family of examples of three-element sinks.

**Theorem 7.4** *Let $\mathbb{A} = (A, F)$ be a three-element idempotent algebra with $A = \{a, b, c\}$ and $s_{abc} \in F$. If all operations in $F$ are $\alpha\beta$-projective, then $\mathbb{A}$ is a sink.*

As just discussed, the operation $s_{abc}$ is $\alpha\beta$-projective, and hence this theorem implies that the algebra $(\{a, b, c\}, \{s_{abc}\})$ is a sink.

**Proof.** Let $\mathbb{A}$ be an algebra of the described form. Observe that $\{a, b\}$ is not a two-element subalgebra of $\mathbb{A}$, since it is not preserved by $s_{abc}$. We show that $\alpha = \{a, c\}$ and $\beta = \{b, c\}$ are two-element subalgebras of $\mathbb{A}$, which implies that $\mathbb{A}$ is fully connected. Let $f : A^k \to A$ be an operation from $F$. Since $f$ is $\alpha\beta$-projective, it can be realized as a projection $g : \{\alpha, \beta\}^k \to \{\alpha, \beta\}$. Observe that $f(\alpha, \ldots, \alpha) \subseteq g(\alpha, \ldots, \alpha) = \alpha$ and $f(\beta, \ldots, \beta) \subseteq g(\beta, \ldots, \beta) = \beta$, implying that $\alpha$ and $\beta$ are both preserved by $f$. We have shown that $\alpha$ and $\beta$ are subalgebras of $\mathbb{A}$.

We now prove that every term operation of $\mathbb{A}$ is $\alpha\beta$-projective. This implies that every term operation of $\mathbb{A}$ can be realized as an operation on $\{\alpha, \beta\}$, which in turn implies that $\mathbb{A}$ is enclosed, as $\alpha$ and $\beta$ are exactly the maximal proper subalgebras of $\mathbb{A}$. First, observe that any projection $f : A^k \to A$ can be realized by the projection $g : \{\alpha, \beta\}^k \to \{\alpha, \beta\}$ that projects onto the same coordinate as $f$. Second, let $f : A^n \to A$ and $f_1, \ldots, f_n : A^m \to A$ be operations that are $\alpha\beta$-projective, and let $g : \{\alpha, \beta\}^n \to \{\alpha, \beta\}$ and $g_1, \ldots, g_n : \{\alpha, \beta\}^m \to \{\alpha, \beta\}$ be projections realizing them, respectively. Then, the composition of $f$ with $f_1, \ldots, f_n$, namely, the operation $f(f_1(x_1, \ldots, x_m), \ldots, f_n(x_1, \ldots, x_m))$ is realized by the corresponding composition $g(g_1(y_1, \ldots, y_m), \ldots, g_n(y_1, \ldots, y_m))$ which is a projection.

We now show that $\mathbb{A}$ does not contain a G-set as factor. First, observe that $\mathbb{A}$ itself is not a G-set, because $s_{abc}$ is not essentially unary. Also, observe that the subalgebra $\alpha$ is not a G-set, because $s_{abc}|_\alpha$ is not essentially unary; likewise, the subalgebra $\beta$ is not a G-set, because $s_{abc}|_\beta$ is not essentially unary. Any other factor of $\mathbb{A}$ of non-trivial size not isomorphic to $\mathbb{A}$, the subalgebra $\alpha$ nor the subalgebra $\beta$ must be a homomorphic image of $\mathbb{A}$ having size two. By Fact 2.16, such a homomorphic image must arise from a congruence $\theta$ having $\alpha$ and $\{b\}$ as its equivalence classes, or a congruence $\theta$ having $\beta$ and $\{a\}$ as its equivalence classes. In either case, the operation $s_{abc}^\theta$ is not essentially unary.

We have shown that $\mathbb{A}$ is fully connected, enclosed, and does not contain a G-set as factor. To establish that $\mathbb{A}$ is a sink, it remains to show that $\mathbb{A}$ is not collapsible. To achieve this, it suffices to demonstrate that $\mathbb{A}$ is not collapsible with source $A$, that is, for any $w \geq 1$, there exists $n \geq 1$ such that the adversary $A^n$ is not $\mathbb{A}$-composable from $\cup_{a \in A}\mathsf{Adv}(n, \{a\}, w, A)$. Fix $w \geq 1$, let $n = w + 1$, and consider an adversary $(A_1, \ldots, A_n)$ such that $(A_1, \ldots, A_n)$ is $\mathbb{A}$-composable from $\cup_{a \in A}\mathsf{Adv}(n, \{a\}, w, A)$. We have $(A_1, \ldots, A_n) \triangleleft f(\mathcal{B}_1, \ldots, \mathcal{B}_k)$ for $\mathcal{B}_1, \ldots, \mathcal{B}_k \in \mathsf{Adv}(n, \{a\}, w, A)$ and $f$ a term operation of $\mathbb{A}$. We have shown that every term operation of $\mathbb{A}$ is $\alpha\beta$-projective, so let us assume that $f$ is realized by the projection $g : \{\alpha, \beta\}^k \to \{\alpha, \beta\}$ which projects onto its $i$th coordinate, where $i \in [k]$. Since $n > w$, there exists a coordinate $j \in [n]$ such that $|\mathcal{B}_{ij}| = 1$. Thus $\mathcal{B}_{ij} \subseteq \alpha$ or $\mathcal{B}_{ij} \subseteq \beta$. From this and the fact that $f$ is realized by $g$, which is the projection onto the $i$th coordinate, we have $f(\mathcal{B}_{1j}, \ldots, \mathcal{B}_{kj}) \subseteq \alpha$ or $f(\mathcal{B}_{1j}, \ldots, \mathcal{B}_{kj}) \subseteq \beta$. From $(A_1, \ldots, A_n) \triangleleft f(\mathcal{B}_1, \ldots, \mathcal{B}_k)$, we have $A_j \subseteq f(\mathcal{B}_{1j}, \ldots, \mathcal{B}_{kj})$ and thus either $A_j \subseteq \alpha$ or $A_j \subseteq \beta$, implying that $A_j \neq A$ and hence $(A_1, \ldots, A_n) \neq A^n$. □

If it could be shown that the problem $\mathsf{QCSP}_c(\Gamma)$ is polynomial-time tractable for any constraint language $\Gamma$ having $s_{abc}$ as a polymorphism, then this result along with Theorem 7.3 would imply a classification of those problems $\mathsf{QCSP}_c(\Gamma)$ over a three-element domain that are polynomial-time tractable. Unfortunately, this is not the case: it is known that there exists a finite constraint language $\Gamma$ over domain $\{a, b, c\}$ having $s_{abc}$ as polymorphism such that $\mathsf{QCSP}_c(\Gamma)$ is coNP-hard [11]. We leave further investigation of the properties and complexity of three-element sinks as an issue for future research.

**Acknowledgements.** The author thanks Manuel Bodirsky, Andrei Bulatov, and Víctor Dalmau for interesting and helpful comments. The author also thanks Argimiro Arratia and Andras Salamon for their suggestions on drafts of this paper.

# References

[1] Bengt Aspvall, Michael F. Plass, and Robert Endre Tarjan. A linear-time algorithm for testing the truth of certain quantified boolean formulas. *Information Processing Letters*, 8(3):121–123, 1979.

[2] E. Böhler, N. Creignou, S. Reith, and H. Vollmer. Playing with boolean blocks, part I: Post's lattice with applications to complexity theory. *ACM SIGACT-Newsletter*, 34(4):38–52, 2003.

[3] F. Börner, A. Bulatov, A. Krokhin, and P. Jeavons. Quantified constraints: Algorithms and complexity. In *Computer Science Logic 2003*, 2003.

[4] Andrei Bulatov. Malt'sev constraints are tractable. Technical Report PRG-RR-02-05, Oxford University, 2002.

[5] Andrei Bulatov. Tractable conservative constraint satisfaction problems. In *Proceedings of 18th IEEE Symposium on Logic in Computer Science (LICS '03)*, pages 321–330, 2003.

[6] Andrei Bulatov and Victor Dalmau. A simple algorithm for mal'tsev constraints. *SIAM Journal of Computing*, 36(1):16–27, 2006.

[7] Andrei Bulatov, Peter Jeavons, and Andrei Krokhin. Classifying the complexity of constraints using finite algebras. *SIAM J. Computing*, 34(3):720–742, 2005.

[8] Andrei Bulatov, Andrei Krokhin, and Peter Jeavons. The complexity of maximal constraint languages. In *ACM Symposium on Theory of Computing*, pages 667–674, 2001.

[9] Andrei A. Bulatov. A dichotomy theorem for constraint satisfaction problems on a 3-element set. *Journal of the ACM (JACM)*, 53, 2006.

[10] Hans Kleine Büning, Marek Karpinski, and Andreas Flögel. Resolution for quantified boolean formulas. *Information and Computation*, 117(1):12–18, 1995.

[11] Hubie Chen. Quantified constraint satisfaction and 2-semilattice polymorphisms. In *CP*, 2004.

[12] Nadia Creignou, Sanjeev Khanna, and Madhu Sudan. *Complexity Classification of Boolean Constraint Satisfaction Problems*. SIAM Monographs on Discrete Mathematics and Applications. Society for Industrial and Applied Mathematics, 2001.

[13] Victor Dalmau. Some dichotomy theorems on constant-free quantified boolean formulas. Technical Report LSI-97-43-R, Llenguatges i Sistemes Informàtics - Universitat Politècnica de Catalunya, 1997.

[14] Victor Dalmau. A new tractable class of constraint satisfaction problems. In *6th International Symposium on Artificial Intelligence and Mathematics*, 2000.

[15] Victor Dalmau. Mal'tsev constraints made simple. ECCC technical report, 2004.

[16] Victor Dalmau. Generalized majority-minority operations are tractable. In *LICS*, 2005.

[17] H.D. Ebbinghaus, J. Flum, and W. Thomas. *Mathematical Logic*. Springer-Verlag, 1984.

[18] Tomás Feder and Moshe Y. Vardi. The computational structure of monotone monadic snp and constraint satisfaction: A study through datalog and group theory. *SIAM J. Comput.*, 28(1):57–104, 1998.

[19] P. Hell and J. Nešetřil. On the complexity of $H$-coloring. *Journal of Combinatorial Theory, Ser.B*, 48:92–110, 1990.

[20] Pavol Hell and Jaroslav Nesetril. *Graphs and Homomorphisms*. Oxford University Press, 2004.

[21] Peter Jeavons. On the algebraic structure of combinatorial problems. *Theoretical Computer Science*, 200:185–204, 1998.

[22] Peter Jeavons, David Cohen, and Martin Cooper. Constraints, consistency, and closure. *Articial Intelligence*, 101(1-2):251–265, 1998.

[23] P.G. Jeavons, D.A. Cohen, and M. Gyssens. Closure properties of constraints. *Journal of the ACM*, 44:527–548, 1997.

[24] Marek Karpinski, Hans Kleine Büning, and Peter H. Schmitt. On the computational complexity of quantified horn clauses. In *CSL 1987*, pages 129–137, 1987.

[25] E. Kiss and M. Valeriote. On tractability and congruence distributivity. In *LICS*, 2006.

[26] Ph.G. Kolaitis and M.Y. Vardi. Conjunctive-query containment and constraint satisfaction. *Journal of Computer and System Sciences*, 61:302–332, 2000.

[27] B. Larose and L. Zádori. Taylor terms, constraint satisfaction and the complexity of polynomial equations over finite algebras. To appear.

[28] Barnaby Martin and Florent Madelaine. Towards a trichotomy for quantified H-coloring. In *Logical Approaches to Computational Barriers, Second Conference on Computability in Europe*, Lecture Notes in Computer Science. Springer, 2006.

[29] R.N. McKenzie, G.F. McNulty, and W.F. Taylor. *Algebras, Lattices and Varieties*, volume I. Wadsworth and Brooks, California, 1987.

[30] Thomas J. Schaefer. The complexity of satisfiability problems. In *Proceedings of the ACM Symposium on Theory of Computing (STOC)*, pages 216–226, 1978.

[31] L. Stockmeyer and A. R. Meyer. Word problems requiring exponential time. In *5th ACM Symp. on Theory of Computing*, pages 1–10, 1973.

[32] A. Szendrei. *Clones in Universal Algebra*, volume 99 of *Seminaires de Mathematiques Superieures*. University of Montreal, 1986.